%% file: main.tex
\documentclass[journal]{IEEEtran}
\IEEEoverridecommandlockouts
\usepackage{cite}
\usepackage{amsmath,amssymb,amsfonts}
\usepackage{mathtools}
\usepackage{algorithmic}
\usepackage{comment}
\usepackage{graphicx}
\usepackage{enumitem}
\usepackage{textcomp}
\usepackage{xcolor}
\usepackage{color,soul}

\usepackage{float}
\def\BibTeX{{\rm B\kern-.05em{\sc i\kern-.025em b}\kern-.08em
    T\kern-.1667em\lower.7ex\hbox{E}\kern-.125emX}}
\usepackage{graphicx}
\usepackage{subcaption}
\usepackage{booktabs}
\usepackage{adjustbox}
\usepackage{placeins}
\usepackage{array}
\usepackage[linesnumbered, ruled, vlined]{algorithm2e}
\usepackage{pgfplots}

\usepackage[hidelinks]{hyperref}
\usepackage[capitalise,noabbrev]{cleveref}

\usepackage{geometry}
\usepackage[acronyms,nonumberlist,nopostdot,nomain,nogroupskip,acronymlists={hidden}]{glossaries}
\newglossary[algh]{hidden}{acrh}{acnh}{Hidden Acronyms}
\glsdisablehyper
\input{acronyms.tex}
\usepackage{tikzpagenodes,etoolbox}
\usetikzlibrary{calc}
\usepackage[contents={}]{background}

\begin{document}

\title{XAInomaly: Explainable and Interpretable Deep Contractive Autoencoder for O-RAN Traffic Anomaly Detection}

\author{\IEEEauthorblockN{Osman Tugay Basaran$^\dagger$, and Falko Dressler}

\IEEEauthorblockA{School of Electrical Engineering and Computer Science, TU Berlin, Germany \\
E-mail: \{basaran, dressler\}@ccs-labs.org
}}
\maketitle
\begin{abstract}
Generative \gls{ai} techniques have become integral part in advancing next generation wireless communication systems by enabling sophisticated data modeling and feature extraction for enhanced network performance.
In the realm of \gls{oran}, characterized by their disaggregated architecture and heterogeneous components from multiple vendors, the deployment of generative models offers significant advantages for network management such as traffic analysis, traffic forecasting and anomaly detection.
However, the complex and dynamic nature of O-RAN introduces challenges that necessitate not only accurate detection mechanisms but also reduced complexity, scalability, and most importantly interpretability to facilitate effective network management.
In this study, we introduce the XAInomaly framework, an explainable and interpretable \gls{ssl} \gls{deepcae} design for anomaly detection in O-RAN.
Our approach leverages the generative modeling capabilities of our SS-DeepCAE model to learn compressed, robust representations of normal network behavior, which captures essential features, enabling the identification of deviations indicative of anomalies.
To address the black-box nature of deep learning models, we propose reactive \gls{xai} technique called fastshap-C, which is providing transparency into the model's decision-making process and highlighting the features contributing to anomaly detection.
\end{abstract}


\glsresetall

\section{Introduction}
\label{sec:intro}

The rapid proliferation of wireless devices and the exponential growth in data demand have necessitated continuous innovation in cellular network architectures.
Traditional \gls{rans} have historically been deployed as monolithic systems, where hardware and software components are tightly integrated and sourced from single vendors.
While this approach has ensured reliability and performance, it has also led to significant limitations in terms of scalability, flexibility, and cost-effectiveness.
The proprietary nature of traditional \gls{rans} hinders interoperability and slows down the adoption of new technologies, making it challenging to meet the evolving requirements of modern wireless communications. To overcome these challenges, the industry has been transitioning towards \gls{oran} architectures.
\gls{oran} introduces a paradigm shift by disaggregating the traditional \gls{rans} components into modular, interoperable units with open interfaces \cite{polese2023understanding,polese2024empowering}.
This disaggregation allows network operators to mix and match components from different vendors, fostering a competitive ecosystem that drives innovation and reduces costs.
\gls{oran} architectures are also characterized by their support for virtualization and software-defined networking (SDN), enabling dynamic resource allocation and more efficient network management.
The complexity and heterogeneity introduced by \gls{oran} architectures, however, present new challenges in network operation and maintenance.
The integration of multi-vendor components and the dynamic reconfiguration of network elements require advanced analytics and automation to ensure optimal performance.
\gls{ai} and \gls{ml} techniques have emerged as key enablers in this context, offering powerful tools for network optimization, self-organization, and intelligent decision-making across different layers of the network \cite{chen2019artificial}. \gls{genai} models, in particular, have shown great potential in modeling complex network behaviors and generating synthetic data for various applications.
Techniques such as \gls{gans} \cite{goodfellow2014generative} and \gls{deepae} can capture intricate patterns in high-dimensional data, making them suitable for tasks like traffic prediction, anomaly detection, and network simulation \cite{goodfellow2016deep}.
By leveraging generative models, network operators can enhance the accuracy of predictive analytics and improve the robustness of network management strategies.
In 5G/6G networks as well as \gls{oran} scenarios \cite{o-ran.towards.6g}, AI and generative models can be deployed across multiple layers, including:
\begin{itemize}[noitemsep, nolistsep]
    \item \textit{Physical Layer:} Enhancing signal processing algorithms, adaptive beamforming, and channel estimation.
    \item \textit{MAC and Network Layers:} Optimizing scheduling, resource allocation, and interference management.
    \item \textit{Application and Service Layers:} Personalizing user experiences, content caching, and service quality prediction.
    \item \textit{Management and Orchestration Layers:} Automating network configuration, fault detection, and performance optimization.
\end{itemize}

Anomalies in \gls{oran} traffic can stem from various sources, including misconfigurations \cite{naula2024misconf}, \gls{ue} failures, software bugs, and security breaches such as cyber-attacks.
Early and accurate detection of these anomalies is crucial to prevent service degradation, ensure user satisfaction, and maintain the reliability of critical communications, especially in the context of 5G+/6G networks that support mission-critical applications.
Traditional supervised learning approaches for anomaly detection rely heavily on large amounts of labeled data, which is often impractical in real-world network environments due to the rarity and unpredictability of anomalies.
Moreover, the manual labeling of network traffic data is labor-intensive and prone to human error.
Therefore, there is a pressing need for models that can effectively detect anomalies with minimal reliance on labeled datasets.

Semi-supervised learning models, particularly \gls{deepae}, offer a compelling solution to this challenge.
However, primary issue of standard \gls{deepae} is focusing solely on reconstructing input data \cite{torabi2023practical} is their tendency to overfit to the training data, capturing both the underlying patterns and the noise present in the data.
Since their objective is purely to minimize reconstruction error, they may not learn robust or meaningful feature representations that generalize well to new, unseen data.
This lack of generalization is particularly problematic in anomaly detection tasks within O-RAN environments, where the data is high-dimensional, dynamic, and contains subtle anomalies.
Also, \gls{deepae}s are often complex models with large number of layers and parameters, which makes them computationally intensive.
This complexity translates to longer training times, higher memory requirements, and increased inference latency, which are challenging for real-time, resource-constrained O-RAN environments in real life deployments.
Training these \gls{deepae}s on high-dimensional O-RAN data involves complex optimization, requiring a large amount of labeled data to prevent overfitting and extensive computational resources. 

\begin{figure}
\centering
\includegraphics[width=0.95\columnwidth]{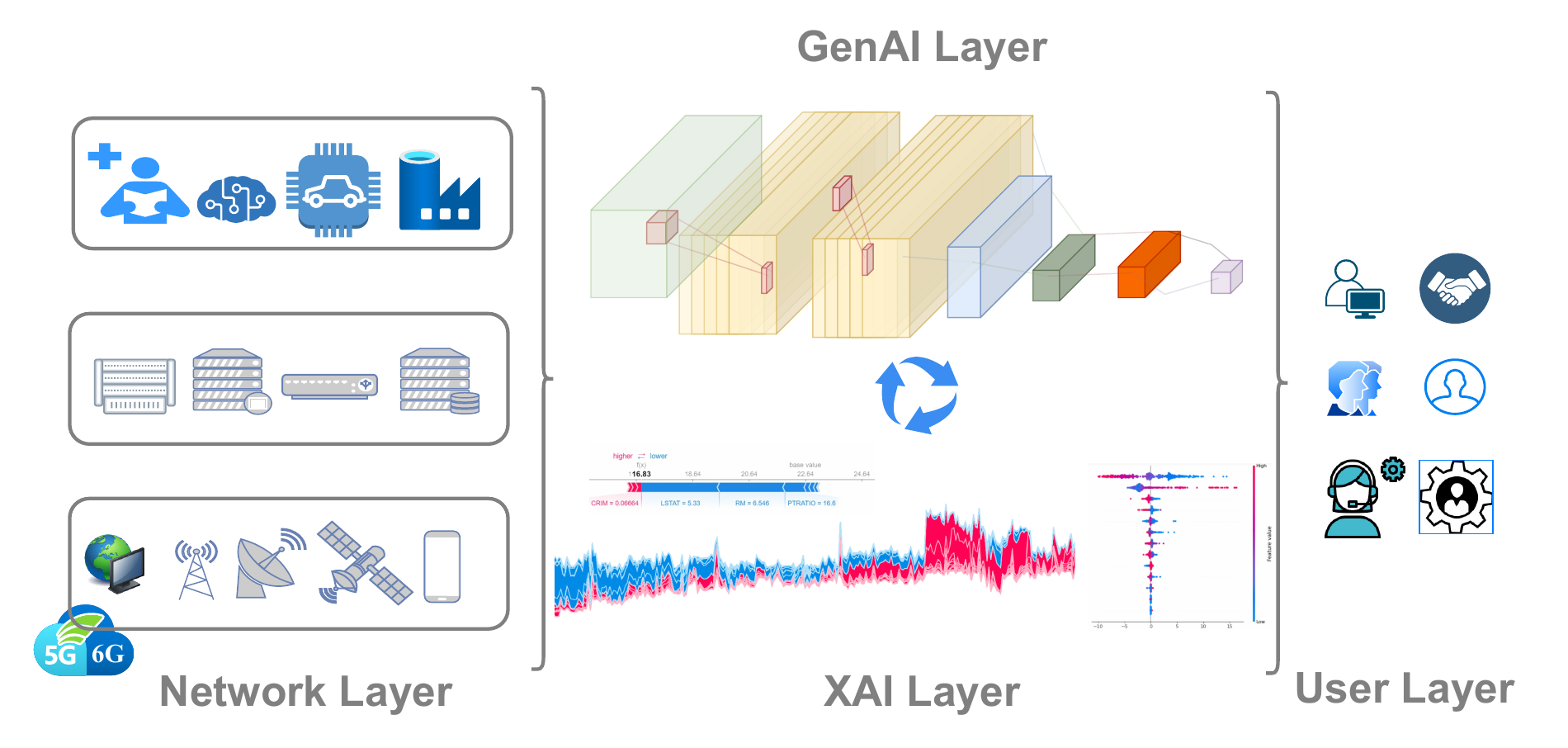}
\caption{GenAI and XAI interaction between user and network layer}
\label{fig:xaiandailayers}
\vspace{-0.55cm}
\end{figure}

Considering the aforementioned challenges, we designed \texttt{SS-DeepCAE} architecture in this study.
\gls{deepcae}s address these drawbacks by introducing a contractive penalty into the loss function.
This penalty encourages the model to learn smooth and robust feature representations by penalizing large sensitivities of the encoder activations with respect to the input.
\gls{deepcae}s tend to have fewer parameters because they do not require complex mechanisms to counter overfitting, such as dropout layers or heavy regularization.
This reduces the model's computational demand, making it more scalable and deployable within resource-constrained O-RAN systems.
Besides, \gls{deepcae}s are better suited for real-time applications due to their reduced sensitivity to noise and more efficient learning of feature representations.
This adaptability is crucial for O-RAN, where anomaly detection may require fast, on-the-fly assessments to maintain network stability and performance.
The lack of existing literature on the use of DCAEs for O-RAN traffic anomaly detection underscores the novelty of this approach.
By filling this gap, we aim to provide a robust framework that leverages the strengths of DCAEs to address the unique challenges posed by O-RAN networks. While autoencoders are powerful, they are often criticized for being “black boxes” due to their opaque internal workings \cite{adadi2018peeking}.
In next generation wireless networks such as 6G \cite{saad2020vision}, this lack of transparency \cite{lipton2018themythos} poses significant risks such as unoptimized network management, poor maintenance, biased monitoring, unfair or suboptimal decisions. To address these concerns, integrating \gls{xai} \cite{gunning2021darpa} techniques into the anomaly detection framework becomes essential  \cite{guo2020explainable}.
As shown in the \Cref{fig:xaiandailayers}, \gls{xai} aims to make \gls{ai} models more transparent by providing understandable explanations of their decisions.
However, integrating \gls{xai} design into \gls{oran} must be done with understanding performance demands of use cases particularly for \gls{urllc}, impose strict constraints on any deployed solutions.
Domain-aware \gls{xai} designs can optimize the trade-off between explainability and computational efficiency by focusing explanations on the most relevant features and events within the \gls{oran} context.
In this direction, we introduced novel \texttt{fastSHAP-C} XAI algorithm in our previous work \cite{basaran2024xainomaly} for xURLLC use case in O-RAN.
However, in this study, we used the \gls{deepae}-based autoencoder that has complexity, scalability and 	generalizability  issues that we mentioned above. Therefore, there is a clear need for an emerging anomaly detection solution that is performance friendly, effective and trustworthy.
In this work, our proposed XAInomaly framework addresses these requirements by combining a novel \texttt{SS-DeepCAE} model with \texttt{fastSHAP-C} XAI algorithm tailored for the dynamic O-RAN traffic environment.

\begin{table*}
\centering
\begin{adjustbox}{width=\textwidth,center}
\begin{tabular}{@{}l|llll|ccccc@{}}
\toprule
Article                                                        & Learning Category                                                               & Learning Model                                                                                                                          & Use Case & Year & Data Availability                                                            & xApps & Explainability                                                                                \\ \midrule
\begin{tabular}[c]{@{}c@{}} Fiandrino et al.
\cite{fiandrino2022toward} \end{tabular}      & Supervised L.
                                                      & LSTM                                                                                                                             & Traffic Management        & 2022  & $\chi$                                                                         & $\chi$                     & \checkmark                                                                                                      \\
\begin{tabular}[c]{@{}c@{}}Mahrez et al.
\cite{mahrez2023benchmarking}\end{tabular}  & Supervised L.
  &  RFO                                                                                                                     & Traffic Anomalies          & 2023   & \checkmark                                                                       & \checkmark                     & $\chi$                                                                                                    \\
\begin{tabular}[c]{@{}c@{}}Alves et al.
\cite{alves2023machine}\end{tabular}  & Supervised L.
  & MLP                                                                                                                  &  Traffic Anomalies          & 2023   & \checkmark                                                                & $\chi$                     & $\chi$                                                                                         \\
\begin{tabular}[c]{@{}c@{}}Fiandrino et al.
\cite{fiandrino2023explora}\end{tabular}      & Supervised L.
  & DRL & Resource Allocation         & 2023 & $\chi$          & \checkmark                   & \checkmark                                                                                                           \\
\begin{tabular}[c]{@{}c@{}}Khan et al.
\cite{khan2024explainable}\end{tabular}        & Supervised L.
                                                        & DRL                                                                                                                              & Resource Management         & 2024 & $\chi$                                                                          & $\chi$                     & \checkmark                                                                                                              \\

\begin{tabular}[c]{@{}c@{}}Tassie et al.
\cite{tassie2024leveraging}\end{tabular}        & Supervised L.
                                                        & CNN                                                                                                                              & Traffic Classification         & 2024 & $\chi$                                                                         & \checkmark                     & \checkmark                                                                                                             \\

\hline
\begin{tabular}[c]{@{}c@{}}Basaran et al.
\cite{basaran2023deep}\end{tabular}  & Semi-Supervised L.
  &  DeepAE                                                                                                                     &  Traffic Anomalies          & 2023   & \checkmark                                                                        & \checkmark                     & $\chi$                                                                                                   \\
\textbf{Our Work}                                                       & \begin{tabular}[c]{@{}l@{}}Semi-Supervised L.\end{tabular}                                                              & DeepCAE                                                                      & Traffic Anomalies       & 2024  & \checkmark                                                                       & \checkmark                    & \checkmark                                                     \\ \bottomrule
\end{tabular}
\end{adjustbox}
\caption{Literature review on different O-RAN use cases with generative and explainable AI}
\label{literature}
  \vspace{-.8em}
\end{table*}

\subsection{Our Contributions}
The core outcomes of our study are summarized as new contributions (“C") and new findings (“F") as follows:

\newcommand\itema{\item[\textbf{C1.}]}
\newcommand\itemb{\item[\textbf{C2.}]}
\newcommand\itemc{\item[\textbf{C3.}]}
\newcommand\itemd{\item[\textbf{F1.}]}
\newcommand\iteme{\item[\textbf{F2.}]}
\newcommand\itemf{\item[\textbf{F3.}]}

\begin{itemize}[noitemsep,nolistsep]
    \itema We propose the XAInomaly framework, which, for the first time, unites \gls{genai} and \gls{xai} design in \gls{oran}.
Our platform provides an interoperable \texttt{SS-DeepCAE} model for the \gls{oran} traffic anomalies.
    \itemb We introduce novel \texttt{fastSHAP-C} XAI method that obtain real-time SHAP values regarding O-RAN intelligence orchestration operations.
    \itemc We are the first to implement scalable, resource-efficient \texttt{SS-DeepCAE} in O-RAN traffic anomaly detection scenario.
    \itemd The ability of \texttt{SS-DeepCAE} to achieve high UAR with minimal labeled data makes it highly scalable for O-RAN environments, where labeled anomalies are rare and costly to obtain.
    \iteme Our results show that \texttt{fastSHAP-C} provides $34\%$ advance over its competitors in terms of runtime performance.
\end{itemize}

The rest of the paper is organized as follows: In \Cref{sec:background} building blocks of \gls{oran} architecture, existing \gls{genai} and \gls{xai} implementations are presented (see \Cref{literature}).
\Cref{sec:autoencoder design} describes the proposed \texttt{SS-DeepCAE} model in detail, including semi-supervised learning approach and architecture.
\Cref{sec:xainomaly} describes proposed XAInomaly framework with selected \gls{xai} interpretation methodology, proposed \texttt{fastSHAP-C} algorithm and XAInomaly integration on \gls{oran} architecture.
\Cref{sec:experiments} includes datasets, performance metrics, experiments with baseline models, and hyper-parameter tuning stages.
\Cref{sec:results}  consists of performance results of the \texttt{SS-DeepCAE} anomaly detection model and proposed novel XAI algorithm \texttt{fastSHAP-C}.
Finally, \Cref{sec:conc} concludes the paper and discusses potential directions for future work with challenges. \Cref{tab:acros} contains important acronyms and definitions used throughout this document.

\begin{table*}
  \centering
  \scalebox{0.93}{
    \begin{tabular}{ll ll}
    \toprule
    Acronym & Definition & Acronym & Definition \\
    \midrule
    3GPP    & 3rd Generation Partnership Project &  O-RU & O-RAN Radio Unit  \\
    5G    & Fifth Generation Mobile Network & PDCP  &  Packet Data Convergence Protocol  \\
    6G    & Sixth Generation Mobile Network & PHY  & Physical \\
    AI    & Artificial Intelligence &  QoE   & Quality of Experience \\
    AD xApp    & Anomaly Detection xApp &   QoS    & Quality of Service \\
    CNN   & Convolutional Neural Network &   RAN  &  Radio Access Networks \\
    $\mathcal{CS}$   & Confidence Score & RF  & Radio Frequency \\
    DL    & Deep Learning & rApps & RAN Applications \\
    DeepAE    & Deep Autoencoder &  RFO & Random Forest\\
    DeepCAE    & Deep Contractive Autoencoder & ReLU & Rectified Linear Unit \\
    DRL    & Deep Reinforcement Learning & RLC & Radio Link Control\\
    $\mathcal{EM}$    & Error Metric   & SDAP  & Service Data Adaptation Protocol  \\
    GenAI    & Generative AI &  SGD & Stochastic Gradient Descent \\
    GANs     & Generative Adversarial Networks  & SHAP & SHapley Additive exPlanations \\
    HARQ    & Hybrid Automatic Repeat reQuest & SLA-VAE & Semi-supervised VAE \\
    LSTM   & Long Short-term Memory & SMO & Service Management and Orchestration  \\
    O-RAN    & Open Radio Access Networks & SS & Semi-supervised   \\
    MAC     & Medium Access Control & SS-DeepCAE & Semi-supervised Deep Contractive Autoen.
 \\
    MSE     & Mean Squared Error & t-SNE & t-Distributed Stochastic Neighbor Embedding  \\
    ML    & Machine Learning & TS xApp & Traffic Steering xApp \\
    Near-RT RIC & Near-Real-Time RAN Intelligent Controller & UE & User Equipment  \\
    Non-RT RIC & Non-Real-Time RAN Intelligent Controller & URLLC & Ultra-Reliable Low-Latency Communications  \\
    O-CU & O-RAN Central Unit &  VAEs & Variational Autoencoders \\
    O-DU & O-RAN Distributed Unit & XAI & Explainable AI \\
    OFH & Open Fronthaul & xApps & eXtended applications \\

    \bottomrule
    \end{tabular}%
  }
  \caption{List of important acronyms and their definitions}
  \label{tab:acros}%
  \vspace{-.8em}
\end{table*}%

\section{Background}
\label{sec:background}

This section provides a detailed overview of the key components of the \gls{oran} architecture, \gls{genai} and \gls{xai} literature on \gls{oran}.

\subsection{Principles of O-RAN}
\label{sec:principoran}

\gls{oran} architecture disaggregates traditional RAN hardware into modular components with standardized interfaces.
O-CU is responsible for the higher layers of the protocol stack, specifically the \gls{pdcp} and \gls{sdap} layers.
\gls{ocu} can be further divided into O-CU-CP (Control Plane) and O-CU-UP (User Plane) for enhanced scalability and flexibility.
\gls{odu} manages the \gls{rlc}, \gls{mac}, and parts of the \gls{phy} layer \cite{o-ran.arch}.
It allocates radio resources to UEs based on scheduling algorithms and handles \gls{harq} for error correction.
\gls{odu} is typically located closer to the radio units to meet stringent latency requirements.
This allows to perform time-sensitive computations required for real-time operations of radio communications.
\gls{oru} encompasses the lower \gls{phy} layer and the \gls{rf} components  \cite{etsi2022}.
\gls{oru} converts digital signals to analog for transmission over the air and vice versa.
It also implements antenna array processing for signal directionality for beamforming.

\gls{oru} interfaces with the O-DU via the \gls{ofh} interface, which is standardized to allow interoperability between different vendors' equipment.
\gls{3gpp} and \gls{oran} Alliance's standardized interfaces are a cornerstone of the \gls{oran} architecture, ensuring interoperability and flexibility.
E2 interface connects the Near-Real-Time RAN Intelligent Controller (Near-RT RIC) with the \gls{rans} components (O-CU and O-DU), enabling real-time control and data exchange \cite{o-ran.E2}.
A1 interface links the \gls{nonrt} with the \gls{nearrt}, facilitating policy management and model updates \cite{o-ran.a1}.
O1 interface connects the \gls{smo} with the \gls{rans} components for management and orchestration tasks \cite{o-ran.01}.
\gls{ofh} interface connects the \gls{odu} with the \gls{oru}, supporting high-bandwidth, low-latency data transfer.

\gls{smo} framework is a central element in the \gls{oran} architecture responsible for the overall management and orchestration of network services.
It provides end-to-end service lifecycle management, including provisioning, configuration, optimization, and assurance of network functions.
It allocates and optimizes physical and virtual resources across the network.
Also manages network slices to support diverse service requirements with different performance and resource needs.
 Hosts AI/ML models and analytics that support intelligent decision-making across the network then implement policies for network operations, security, and compliance.

\gls{nonrt} operates within the \gls{smo} framework and provides non-real-time control and optimization of \gls{rans} functions \cite{o-ran.smo}.
It leverages AI/ML algorithms to perform tasks that do not require immediate responsiveness, typically with latency requirements greater than one second.
It generates and updates policies for \gls{rans} optimization, which are communicated to the \gls{nearrt} via the A1 interface.
Performs extensive data collection and analysis to understand network performance afterwards it develops and refines AI/ML models based on long-term data analysis.
\gls{nearrt} is a critical component responsible for real-time and near-real-time control of \gls{rans} elements, with latency requirements typically between 10 ms and 1 s.
It enables dynamic optimization of \gls{rans} resources to meet immediate performance objectives.
It adjusts parameters such as scheduling, power control, and beamforming in time sensitive manner.

\begin{figure}
    \includegraphics[width=\columnwidth]{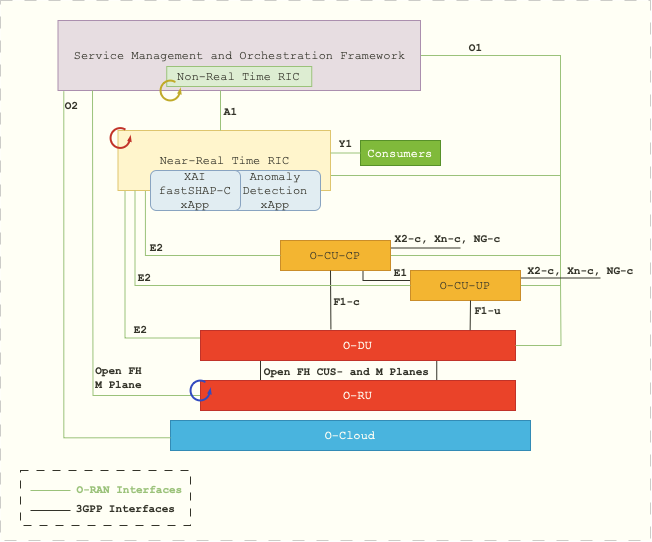}
    \caption{O-RAN reference architecture with XAInomaly}
    \label{oranarch}
    \vspace{-.8em}
\end{figure}

\gls{rans} applications are software modules that provide specialized functionalities within the \gls{oran} architecture.
They are categorized into \gls{rapps} and \gls{xapps} based on their operational time scales and hosting environments.
\gls{rapps} are \gls{rans} applications that run on the \gls{nonrt} within the \gls{smo} framework.
They perform non-real-time functions such as network planning with long-term optimization of network topology and configurations, forecasting network behavior and traffic patterns, automated adjustment of network settings based on policies and analytics.
As shown in \Cref{oranarch}, our proposed \gls{xapps} hosted on the \gls{nearrt}, designed for real-time and near-real-time operations \cite{o-ran.near}.
They are useful for dynamic distribution of traffic to optimize network utilization, immediate adjustments to meet \gls{qos} requirements for different services.

\subsection{GenAI and XAI Literature on O-RAN}
\label{sec:genxaiporan}

\gls{genai} studies on wireless communication systems are quite common in the literature \cite{fang2024prism, zeng2023cvca, zeng2024csi}.
\gls{genai} models, such as \gls{gans}, \gls{vaes}, and \gls{deepae}s, have demonstrated significant capabilities in modeling complex data distributions and generating synthetic data.
Similarly, \gls{genai} studies on O-RAN are quite numerous however, XAI research is still scarce, but gaining momentum recently.
In \Cref{literature}, we gathered and summarized the current \gls{genai} and \gls{xai} research on \gls{oran}.
Studies focused on resource allocation, traffic anomalies and management use cases.

In resource allocation studies, Khan \textit{et al.} \cite{khan2024explainable} explores the application of XAI within distinct 6G use cases, with a focus on a \gls{drl} solution designed for vehicular network scenarios.
Specifically, the authors utilized SHAP values to analyze feature contributions.
However, a major challenge lies in the high complexity and dimensionality of \gls{drl} models, which often involve many layers and parameters, making interpretation difficult.
This complexity can result in explanations that either oversimplify and miss important details or are too intricate for end-users to easily grasp.
Moreover, in a vehicular use case, the XAI model must provide real-time evaluations, a requirement that conventional SHAP algorithms struggle to meet in practical settings due to computational demands.
In \cite{fiandrino2023explora}, the authors present EXPLORA, an innovative framework designed to improve the explainability of AI/ML models within Open RAN systems.
This study highlights the critical need for transparency and interpretability in decentralized networks like Open RAN, where complex decisions are often made at the network edge.

Different models have been designed specifically for traffic management studies.
When the studies are examined, it is seen that most of the models are supervised learning based solutions such as \gls{rfo}, Multilayer Perceptron (MLP), Convolutional Neural Network (CNN) \cite{mahrez2023benchmarking, alves2023machine, tassie2024leveraging}.However, considering heterogeneous components of O-RAN, especially the specific interfaces and data flows, extensive data labeling is quite costly and challenging.
Fiandrino \textit{et al.} \cite{fiandrino2022toward} further investigate the application of XAI and robustness in 6G networks, acknowledging the challenges that remain, particularly in achieving a balance between explainability and robustness while providing explanations that are both precise and accessible to diverse stakeholders, such as network operators and end-users.
A use case involving a 4G network examines the LRP and SHAP algorithms, with shared results on execution time and resource usage.
The findings reveal that CPU usage poses a significant limitation, indicating the need for the development of more advanced, domain-specific XAI algorithms.
This motivated our research into creating XAI solutions capable of managing the real-time data demands of 6G xURLLC communications.

\section{Semi-supervised DeepCAE Design for O-RAN Anomaly Detection}
\label{sec:autoencoder design}
\subsection{Overview of Our Semi-Supervised L.
Approach}

In the context of anomaly detection within O-RAN networks, semi-supervised learning is particularly advantageous due to the scarcity of labeled anomalous data.
Semi-supervised learning leverages both labeled and unlabeled data during training, improving the model's ability to generalize and detect unseen anomalies~\cite{zhu2009introduction}.
In our approach, the majority of the training data consists of unlabeled normal traffic patterns, while a small subset is labeled to guide the learning process.
Given the labeled dataset
\begin{equation}
    \mathcal{D}_l = \{ (\mathbf{x}_i, y_i) \}_{i=1}^{N_l}
\end{equation}
where $\mathbf{x}_i \in \mathbb{R}^n$ represents the input features, and $y_i \in \{0,1\}$ is the label indicating normal ($0$) or anomalous ($1$) traffic.
Given also the unlabeled dataset
\begin{equation}
    \mathcal{D}_u = \{ \mathbf{x}_i \}_{i=N_l+1}^{N_l+N_u}
\end{equation}
where $N_u \gg N_l$, reflecting the abundance of unlabeled data.
The total dataset is
\begin{equation}
\mathcal{D} = \mathcal{D}_l \cup \mathcal{D}_u
\end{equation}
with total samples $N = N_l + N_u$.

Our objective is to train a model that can
\begin{enumerate}[noitemsep, nolistsep, label=\large\protect\textcircled{\small\arabic*}]
    \item Reconstruct normal traffic patterns accurately.
    \item Learn robust and contractive latent representations to detect anomalies.
    \item Incorporate very few label information to improve anomaly detection performance.
    \item Generalize well to detect unseen anomalies.
\end{enumerate}

\subsection{Proposed DeepCAE Architecture}

Our \gls{deepcae} introduces a regularization term that penalizes the sensitivity of the encoder's output to small perturbations in the input ~\cite{hinton2006reducing}.
This encourages the model to learn robust representations that capture
essential features.
Architecture consists of two main components
\begin{enumerate}
    \item \textit{Encoder ($E$)}: Transforms the input data $\mathbf{x} \in \mathbb{R}^n$ into a latent representation $\mathbf{z} \in \mathbb{R}^m$, where $m < n$.
    \item \textit{Decoder ($D$)}: Reconstructs the input data from the latent representation, producing $\hat{\mathbf{x}} \in \mathbb{R}^n$.
\end{enumerate}

DeepCAE is trained to minimize the reconstruction error between $\mathbf{x}$ and $\hat{\mathbf{x}}$, effectively learning the underlying structure of the data.
The encoder $E$ maps input data $\mathbf{x}$ to a latent representation $\mathbf{z}$
\begin{equation}
\mathbf{z} = E(\mathbf{x}; \theta_e)=f_e(\mathbf{W}_e \mathbf{x} + \mathbf{b}_e)
\end{equation}
where $\theta_e = \{ \mathbf{W}_e, \mathbf{b}_e \}$ represents the encoder parameters.
In layer-wise form, the encoder can be expressed as
\begin{equation}
\mathbf{h}^{(l)} = f^{(l)}(\mathbf{W}^{(l)} \mathbf{h}^{(l-1)} + \mathbf{b}^{(l)})
\end{equation}
for $l = 1, 2, \dots, L_e$, where: $\mathbf{h}^{(0)} = \mathbf{x}$, $\mathbf{h}^{(L_e)} = \mathbf{z}$, $\mathbf{W}^{(l)}$ and $\mathbf{b}^{(l)}$ are weights and biases for layer $l$, $f^{(l)}(\cdot)$ is the activation function.
\footnote{Activation function selected as Rectified Linear Unit (ReLU) because ReLU inherently induces sparsity by outputting zero for any negative input, which encourages the model to learn sparse representations.
Sparse features are often more effective for distinguishing normal data from anomalies, as they can emphasize essential patterns while suppressing noise.
Also, ReLU mitigates the vanishing gradient problem common with saturating activations (e.g., sigmoid or tanh), which allows for deeper networks to learn effectively.
In anomaly detection, deeper layers are beneficial for capturing complex patterns and subtle distinctions between normal and anomalous data.} Decoder $D$ reconstructs the input from $\mathbf{z}$:
\begin{equation}
\hat{\mathbf{x}} = D(\mathbf{z}; \theta_d) = f_d(\mathbf{W}_d \mathbf{z} + \mathbf{b}_d)
\end{equation}
where  $\theta_d = \{ \mathbf{W}_d, \mathbf{b}_d \}$ represents the decoder parameters.
Similarly, in layer-wise form:
\begin{equation}
\mathbf{h}^{(l)} = f^{(l)}(\mathbf{W}^{(l)} \mathbf{h}^{(l-1)} + \mathbf{b}^{(l)})
\end{equation}
for $l = L_e+1, L_e+2, \dots, L$, with: $\mathbf{h}^{(L_e)} = \mathbf{z}$, $\mathbf{h}^{(L)} = \hat{\mathbf{x}}$.

\subsubsection{Loss Functions}
For all samples (both labeled and unlabeled), we compute the reconstruction loss to train the autoencoder to accurately reconstruct normal input patterns:
\begin{equation}
\ell_{\text{recon}}(\mathbf{x}_i, \hat{\mathbf{x}}_i) = \| \mathbf{x}_i - \hat{\mathbf{x}}_i \|_2^2
\end{equation}

To encourage robustness in the latent representation, we introduce a contractive penalty based on the Frobenius norm of the Jacobian of the encoder activations with respect to the input \cite{rifai2011contractive}:
\begin{equation}
\ell_{\text{contractive}}(\mathbf{x}_i) = \lambda_c \left\| \frac{\partial E(\mathbf{x}_i; \theta_e)}{\partial \mathbf{x}_i} \right\|_F^2
\end{equation}
where:  $\lambda_c$ is the regularization parameter controlling the strength of the penalty.
$\left\| \cdot \right\|_F^2$ denotes the squared Frobenius norm.
For the deep encoder, the Jacobian $\mathbf{J}_i$ is:
\begin{equation}
\begin{multlined}
    \mathbf{J}_i = \\
    \frac{\partial \mathbf{z}_i}{\partial \mathbf{x}_i} = \mathbf{W}^{(l)} \text{diag}\left( f'^{(l-1)}(\mathbf{W}^{(l-1)} \mathbf{x}_i + \mathbf{b}^{(l-1)}) \right) \mathbf{W}^{(l-1)}
\end{multlined}
\end{equation}
where $f'^{(l-1)}(\cdot)$ is the derivative of the activation function in the first layer.
For very few labeled samples, we introduce a cross-entropy loss\footnote{Typically, a sigmoid function is used in the output layer, which can lead to a saturation effect in the loss function, causing it to plateau.
This saturation hampers gradient-based learning algorithms, limiting their ability to make progress.
To mitigate this issue, incorporating a logarithmic term in the objective function counteracts the exponential nature of the sigmoid function, facilitating effective gradient flow.
Consequently, binary cross-entropy loss is preferred over \gls{mse} since it incorporates a logarithmic term, unlike \gls{mse} \cite{goodfellow2016deep}.} to incorporate the label information into the training process.
We first map the latent representation $\mathbf{z}$ to a prediction $\hat{y}$
\begin{equation}
\ell_{\text{cro}}(y_i, \hat{y}_i) = -[ y_i \log \hat{y}_i + (1 - y_i) \log (1 - \hat{y}_i) ]
\end{equation}

Then total loss function combines the reconstruction loss and supervised loss
\begin{equation}
\begin{multlined}
    L = \frac{1}{N} \sum_{i=1}^{N} [ \ell_{\text{recon}}(\mathbf{x}_i, \hat{\mathbf{x}}_i) + \ell_{\text{contractive}}(\mathbf{x}_i) + \\  \alpha_i \cdot \ell_{\text{cro}}(y_i, \hat{y}_i) ]
\end{multlined}
\end{equation}
where
\begin{equation}
\alpha_i = \begin{cases}
\lambda, & \text{if } i \leq N_l \\
0, & \text{if } i > N_l
\end{cases}
\end{equation}
and $\lambda$ is a hyper-parameter controlling the influence of the cross-entropy loss.

\subsubsection{Optimization and Batch Training}
We optimize the combined parameters $\theta = \{ \theta_e, \theta_d\}$ by minimizing the total loss $L$ using gradient-based optimization methods, such as Adam or SGD with momentum.
\footnote{Although we do not discuss here, we also remark that the Adam optimizer is selected \cite{kingma2014adam}.} The update rule is:
\begin{equation}
\theta \leftarrow \theta - \eta \nabla_\theta L
\end{equation}
where $\eta$ is the learning rate.
The gradient of the contractive penalty requires computing the derivative of the Frobenius norm:
\begin{equation}
\nabla_{\theta_e} \ell_{\text{contractive}}(\mathbf{x}_i) = 2 \lambda_c \left( \frac{\partial E(\mathbf{x}_i; \theta_e)}{\partial \mathbf{x}_i} \right)^\top \frac{\partial^2 E(\mathbf{x}_i; \theta_e)}{\partial \mathbf{x}_i \partial \theta_e}
\end{equation}

Efficient computation techniques, automatic differentiation, are employed to compute these gradients.
During training, we use mini-batches that contain both labeled and unlabeled samples.
Each batch is constructed by sampling $B_l$ labeled and $B_u$ unlabeled samples, ensuring that the model learns from both types of data in each iteration.

\subsubsection{Anomaly Detection During Inference}
After training, we use the model to detect anomalies in new data.
For an input $\mathbf{x}$, we compute the reconstruction error:
\begin{equation}
\text{RE}(\mathbf{x}) = \| \mathbf{x} - \hat{\mathbf{x}} \|_2^2
\end{equation}

\noindent If $\text{RE}(\mathbf{x})$ exceeds a threshold $\tau$, we classify $\mathbf{x}$ as anomalous:
\begin{equation}
\text{Anomaly Indicator} = \begin{cases}
1, & \text{if } \text{RE}(\mathbf{x}) > \tau \\
0, & \text{otherwise}
\end{cases}
\end{equation}

\noindent Anomalies may result in larger norms of the latent representation due to the contractive penalty:
\begin{equation}
\text{Latent Norm} = \| \mathbf{z} \|_2
\end{equation}

\noindent Therefore, we can define an anomaly score combining reconstruction error and latent norm:
\begin{equation}
\text{Anomaly Score} = \gamma \cdot \text{RE}(\mathbf{x}) + (1 - \gamma) \cdot \| \mathbf{z} \|_2
\end{equation}

\noindent where $\gamma \in [0,1]$ balances the contributions.
Model classifies $\mathbf{x}$ as anomalous if the Anomaly Score exceeds a threshold.
 The latent representation $\mathbf{z}$ provides a compressed embedding:
\begin{equation}
m = \dim(\mathbf{z}) < n = \dim(\mathbf{x})
\end{equation}

\noindent Our contractive penalty ensures that $\mathbf{z}$ is less sensitive to small input variations, enhancing robustness.

%

\section{Proposed Design: XAInomaly Framework}
\label{sec:xainomaly}

In this section, we introduce our \texttt{fastSHAP-C} algorithm, an efficient and interpretable method designed to explain the predictions of our DeepCAE used for anomaly detection in O-RAN networks.
The algorithm extends the existing fastSHAP method by incorporating a Confidence Score $\mathcal{CS}$ and an Error Metric $\mathcal{EM}$ to assess the reliability and accuracy of the explanations, which is critical for 5G+/6G O-RAN applications.
\subsection{Selecting XAI Interpretation Methods}
In our implementation, we have designed \texttt{fastSHAP-C} algorithm with Global Model-Agnostic, and Reactive \gls{xai} perspective.

\subsubsection{Global Model-Agnostic XAI Interpretation}
Global interperation provide explanations that encompass the overall behavior of our model across the entire dataset.
This reduces computational overhead, allowing for efficient processing of vast amounts of data and ensuring that the \gls{xai} system remains performant as the network complexity grows.

Let \( f: \mathbb{R}^d \rightarrow \mathbb{R} \) be our DeepCAE model, where \( d \) is the number of features.
The goal of a global explanation is to compute the expected contribution of each feature \( i \) to the model's output over the data distribution \( \mathcal{D}(x) \)
\begin{equation}
\phi_i^{\text{global}} = \mathbb{E}_{x \sim \mathcal{D}(x)} \left[ \phi_i(x) \right]
\label{eq:global_shapley_value}
\end{equation}
where \( \phi_i(x) \) is the Shapley value of feature \( i \) for input \( x \), representing the contribution of feature \( i \) to the prediction at \( x \).
In \texttt{fastSHAP-C}, we approximate the global Shapley values by training an explainer function \( \phi_{\text{fast}}(x;\theta) \) parameterized by \( \theta \) to predict \( \phi_i(x) \) for any input \( x \)
\begin{equation}
\phi_i(x) \approx \phi_{\text{fast},i}(x;\theta)
\label{eq:fastshap_approximation}
\end{equation}

By learning \( \theta \) over the entire dataset, we ensure that the explainer captures the global behavior of the model.
Model-agnostic \gls{xai} methods do not rely on the internal structure or parameters of the predictive model.
Instead, they use input-output behavior to generate explanations.
Our \texttt{fastSHAP-C} algorithm treats \gls{deepcae} model \( f(x) \) as a black box.
The Shapley values are computed based on the model's predictions for different subsets of features without requiring access to the model's internals.
The value function \( v(S) \) for a subset of features \( S \subseteq \{1, 2, \dots, d\} \) is defined as
\begin{equation}
v(S) = \mathbb{E}_{x' \sim \mathcal{D}(x)} \left[ f(x_{S} \cup x'_{\bar{S}}) \right]
\label{eq:value_function}
\end{equation}
where
\begin{itemize}
    \item \( x_{S} \) is the vector containing the values of features in \( S \) from \( x \).
    \item \( x'_{\bar{S}} \) is a sample from the background distribution for features not in \( S \).
    \item \( \bar{S} \) denotes the complement of \( S \).
\end{itemize}

Shapley value \( \phi_i(x) \) for feature \( i \) is computed as
\begin{equation}
\phi_i(x) = \sum_{S \subseteq N \setminus \{i\}} \frac{|S|!(d - |S| - 1)!}{d!} \left( v(S \cup \{i\}) - v(S) \right)
\label{eq:shapley_value}
\end{equation}
This formulation does not depend on the model's architecture, making \texttt{fastSHAP-C} model-agnostic.

\subsubsection{Reactive XAI Interpretation}
Reactive \gls{xai} methods generate explanations simultaneously with the model's predictions, providing immediate insights.
In the dynamic and time-sensitive context of O-RAN networks, real-time explanations enable prompt understanding and response to anomalies.
It supports immediate decision-making processes, such as triggering mitigation strategies or adjusting network parameters.
In \texttt{fastSHAP-C}, the explainer \( \phi_{\text{fast}}(x;\theta) \) is designed to produce Shapley value estimates rapidly for any input \( x \) by leveraging the pre-trained parameters \( \theta \)
\begin{equation}
\phi(x) = \phi_{\text{fast}}(x;\theta)
\label{eq:reactive_explanation}
\end{equation}
Since \( \phi_{\text{fast}} \) is a learned function, computing \( \phi(x) \) involves a simple forward pass, allowing for real-time explanations.

\SetKwComment{Comment}{/* }{ */}

\begin{algorithm}[t]
\caption{\texttt{fastSHAP-C} Training}\label{alg:fastshap-c}
\textbf{Input:} Value function $f_{x,y}$, learning rate $\alpha$\\
\textbf{Output:} $fastSHAP$ explainer $\phi_{fast}(x,y;\theta) $ \\
initialize $\phi_{fast}(x,y;\theta)$ \\
\While{$not \hspace{0.2cm}  converged$}{
    $sample \hspace{0.2cm} x \sim p(x), y \sim Unif(y), s \sim p(s)$ \\
    $predict \hspace{0.2cm} \hat\phi \gets \phi_{fast} (x,y;\theta)$ \\
    \If{$normalize$} {
        $set \hspace{0.2cm} \hat\phi \gets \hat\phi+d^{-1}(f_{x,y}(1)-f_{x,y}(0)-1^{T}\hat\phi) $ \\
    }
    $calculate$\\
    $\mathcal{L} \gets (f_{x,y}(s)-f_{x,y}(0)-s^{T}\hat\phi)^{2}$\\
    $update \hspace{0.2cm} \theta \gets \theta - \alpha\nabla_\theta\mathcal{L} $
}
$\mathcal{CS} \leftarrow \frac{1}{n} \sum_{i=1}^n \left| f_{x,y}(s_i) - f_{x,y}(0) - s_i^T \hat{\phi} \right|$ \\
$\mathcal{EM} \leftarrow \frac{1}{n} \sum_{i=1}^n \left( f_{x,y}(s_i) - s_i^T \hat{\phi} \right)^2$ \\
\textbf{return} $\phi_{\text{fast}}(x,y;\theta)$, $\mathcal{CS}$, $\mathcal{EM}$ \vspace{0.3cm}
\end{algorithm}

\subsection{Proposed fastSHAP-C Algorithm}
\texttt{fastSHAP-C} algorithm aims to approximate the Shapley values for feature attribution in a computationally efficient manner.
Traditional SHAP methods can be computationally intensive, especially for models with high-dimensional input data, as is common in O-RAN traffic analysis.
\texttt{fastSHAP-C} addresses this challenge by learning a universal explainer that approximates Shapley values for similar data points without needing to optimize separately for each input and incorporating Confidence Score ($\mathcal{CS}$) and Error Metric ($\mathcal{EM}$) to quantify the reliability and accuracy of the explanations.
Inputs and steps of \texttt{fastSHAP-C} algorithm (cf. \Cref{alg:fastshap-c}) are as follows:
\begin{enumerate}[noitemsep, nolistsep, label=\large\protect\textcircled{\small\arabic*}]
    \item \textit{Model} $f_{x,y}$: The value function representing the prediction model (autoencoder) for input $x$ and output $y$.
    \item \textit{Learning Rate} $\alpha$: The step size used in the optimization algorithm.
    \item \textit{Data Sample} $x$: The input data point to be explained.
    \item \textit{Background Dataset} $X$: A set of samples representing the background data distribution.
    \item \textit{Number of Samples} $M$: The number of subsets sampled for approximation.
\end{enumerate}

\begin{enumerate}
\item \textit{Initialize:}
\begin{align*}
    \phi_i(x) & = 0 \quad \forall i \\
    f(x) & = f(x) \\
    \hat{f}(x) & = 0
\end{align*}

\item \textit{Sampling:}
For each sample $m$ from 1 to $M$:
Sample a subset $S \subseteq \{1, 2, \ldots, d\}$ uniformly at random.

\item \textit{Marginal Contribution:}
For each feature $i$:
Compute the marginal contribution of feature $i$ given the subset $S$:
\begin{equation}
    \Delta f_i(S) = f(S \cup \{i\}) - f(S)
\end{equation}

\textit{Update the Shapley value:}
\begin{equation}
    \phi_i(x) = \phi_i(x) + \frac{\Delta f_i(S)}{M}
\end{equation}

\item \textit{Calculate $\mathcal{CS}$:} \\
First compute the reconstructed prediction;
\begin{equation}
    \hat{f}(x) = \sum_{i=1}^d \phi_i(x)
\end{equation}

\begin{equation}
    \mathcal{CS}  = \frac{1}{n} \sum_{k=1}^n \left| f(x_k) - \hat{f}(x_k) \right|
\end{equation}

\noindent The $\mathcal{CS}$  quantifies the average absolute discrepancy between the true model output difference and the explainer's approximation over $n$ samples.
A lower $\mathcal{CS}$  indicates higher confidence in the explanations.

\item \textit{Calculate $\mathcal{EM}$:}
Compute the error between the actual and reconstructed predictions:
\begin{equation}
    EM = \frac{1}{n} \sum_{k=1}^n \left( f(x_k) - \hat{f}(x_k) \right)^2
\end{equation}

\noindent  $\mathcal{EM}$ measures the mean squared error between the true model output and the explainer's approximation.
 A lower $\mathcal{EM}$ signifies higher accuracy of the explainer.

\item \textit{Output:}
Return the Shapley values $\phi_i(x)$, $\mathcal{CS}$, and $\mathcal{EM}$.
\end{enumerate}

The loss function $\mathcal{L}$ measures the discrepancy between the true model output and the approximation provided by the explainer
\begin{equation}
    \mathcal{L} = \left(f_{x,y}(s) - f_{x,y}(\mathbf{0}) - s^\top \hat{\phi}\right)^2
\end{equation}
where
\begin{enumerate}[noitemsep, nolistsep, label=\large\protect\textcircled{\small\arabic*}]

    \item $f_{x,y}(s)$: The model output when features in subset $s$ are present.
    \item $f_{x,y}(\mathbf{0})$: The model output when no features are present (baseline prediction).
    \item $s^\top \hat{\phi}$: The linear combination of predicted Shapley values corresponding to the subset $s$.
\end{enumerate}
Using gradient descent, we update the explainer parameters $\theta$
\begin{equation}
    \theta \leftarrow \theta - \alpha \nabla_\theta \mathcal{L}
\end{equation}

\subsection{XAInomaly Integration on O-RAN}
\label{sec:xainomalyinteg}

\begin{figure}
    \includegraphics[width=\columnwidth]{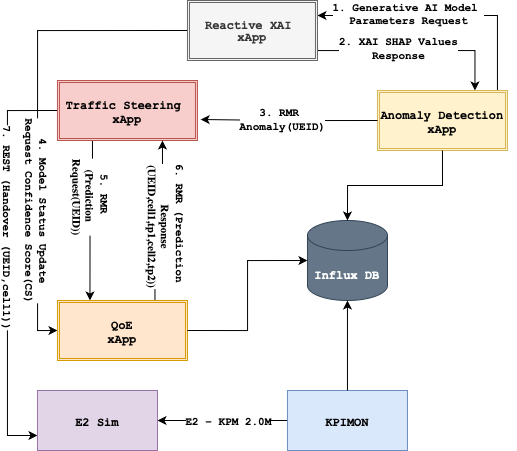}
    \caption{Integration of XAInomaly framework to O-RAN}
    \label{oranflow}
    \vspace{-.8em}
\end{figure}

Our XAInomaly framework is integrated into the O-RAN architecture via \gls{adxapp} and Reactive-XAI xApp as seen in the \Cref{oranflow}.
\gls{adxapp}is central to the anomaly detection process.
It continuously monitors \gls{ue} data, detects anomalies using our \texttt{SS=DeepCAE} model, and identifies UEs exhibiting unusual traffic behavior, degraded performance.
Once an anomaly is detected, the \gls{adxapp} logs this information to the InfluxDB database for record-keeping and further analysis.
It then sends the anomalous \gls{ue} information to the \gls{tsxapp} via Reliable Message Routing (RMR).
\gls{adxapp} also interacts with the Reactive XAI xApp by responding to model parameter requests and to calculate SHAP values, which enable explainability for the detected anomalies.

\gls{tsxapp} is responsible for managing and optimizing \gls{ue} traffic based on detected anomalies.
Upon receiving the list of anomalous UEs from the AD xApp, it requests predictions from the \gls{qoe} Predictor xApp to determine potential throughput for each UE.
\gls{tsxapp}  requests $\mathcal{CS}$ for the anomaly detection process from the Reactive XAI xApp.
This score, calculated by fastSHAP-C, helps gauge the reliability of the anomaly detection, allowing \gls{tsxapp} to factor in the confidence level of detected anomalies during handover decisions.
Based on the throughput predictions provided by the QoE Predictor
xApp, \gls{tsxapp} decides whether to hand over a \gls{ue} to a neighboring cell with better performance metrics (higher predicted throughput).
\gls{tsxapp} works collaboratively with both the \gls{adxapp} and the QoE xApp to make intelligent traffic management decisions and improve network efficiency.


\gls{oran} Software Community's E-release of the \gls{nearrt}\footnote{OSC Near Realtime RIC, \url{ https://wiki.o-ran-sc.org/display/RICP/2022-05-24+ Release+E}} and E2 Agent has been integrated into our XAInomaly framework to manage E2 Service Model (E2SM) functionalities, facilitating communication between the \gls{nearrt} and E2 nodes.
E2SM-KPM is used to gather and report essential KPIs and measurements from the RAN to the \gls{nearrt}.
This continuous flow of real-time metrics enables the RIC to maintain a close watch on network performance.
Anomaly detection and optimization decisions are then enacted through the E2SM-RC (RAN Control) model, which transmits control policies back to the RAN.
By combining E2SM-KPM and E2SM-RC for policy implementation, this setup empowers the \gls{nearrt} to make data-driven decisions that optimize network performance in real-time.
E2 Simulator provides simulated RAN data through E2 (E2-KPM 2.0M) messages, enabling the testing of the xApps.
It supplies critical information on key performance metrics (KPIs) to the xApps.


%

\section{Experimental Analysis}
\label{sec:experiments}

\subsection{Dataset}
\label{subsec:data}

\begin{figure}
\centering
\includegraphics[width=\columnwidth]{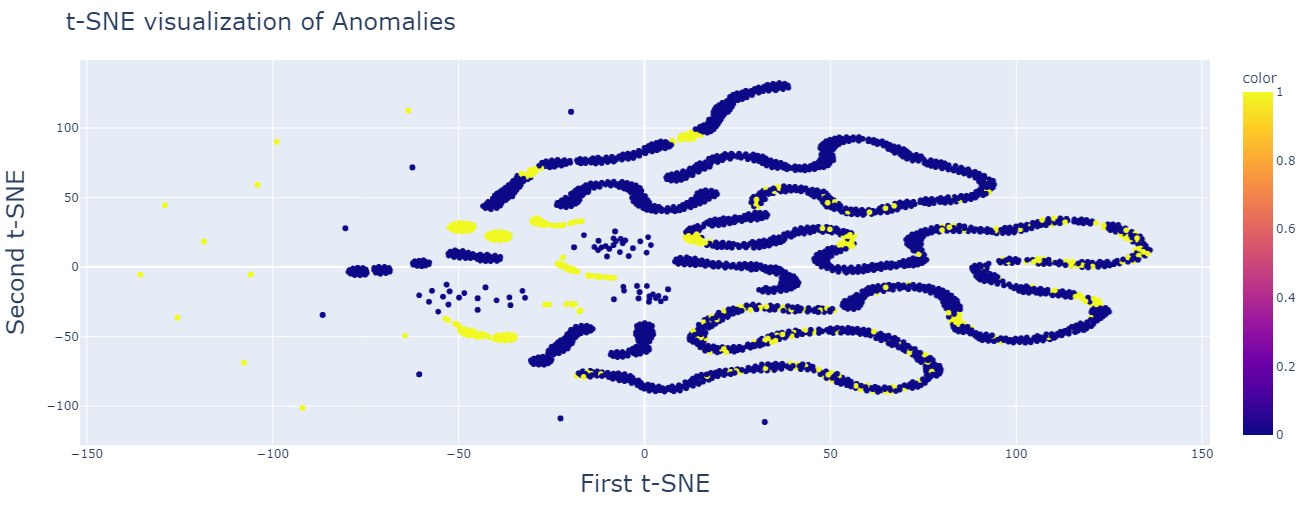}
\caption{Dimensionality reduction on our high-dimensional unbalanced data (yellow color represent anomalous samples, blue color represents non-anomalous samples)}
\label{fig:dataset}
\vspace{-.8em}
\end{figure}

\gls{oran} dataset\footnote{Access to the documented data by \gls{oran} Alliance is provided at \url{https://github.com/o-ran-sc/ric-app-ad/blob/master/src/ue.csv}} was collected from various UE sources, including cars, train passengers, pedestrians, and people waiting.
It comprises 10,000 samples that capture important features such as RSRP, RSRQ, RSSINR, Physical Resource Block (PRB) usage, and Throughput.
Additionally, it includes contextual features like timestamp, NRCellIdentity, DU-id, and UE-id.
The dataset also has a binary label (0 or 1) indicating the presence or absence of an anomaly, with approximately 25\% of the samples labeled as anomalies, highlighting an imbalanced dataset.
In \Cref{fig:dataset}, the dataset's classes are visualized using the \gls{tsne} technique \cite{van-der-maaten2008visualizing}, which reduces high-dimensional data into a two-dimensional view.
The figure clearly shows how anomaly samples are scattered and sparsely located, underscoring the need for a robust model to effectively classify this minority class.

\subsection{Selected Anomaly Detection Metrics for \texttt{SS-DeepCAE}}

We employed six standard metrics to assess the efficacy of anomaly detection \cite{bashar2020regularising}: Accuracy ($ACC$), representing the ratio of correctly classified instances; Precision ($PR$), indicating the ratio of positive instances that are accurately identified as positive; Recall ($RE$), which reflects the ratio of actual positive instances that are correctly recognized; F1 Score ($F_1$), defined as the harmonic mean of precision and recall; Area Under Curve ($AUC$), which gauges the model's ability to differentiate between positive and negative instances and Unweighted Average Recall ($UAR$) also known as Unweighted Average Accuracy, is the sum of class-wise accuracy (recall) divided by number of classes.

\subsection{Selected XAI Metrics for \texttt{fastSHAP-C}}

To enhance the interpretability and trustworthiness of our \texttt{SS-DeepCAE} for anomaly detection in \gls{oran} networks, we integrate specific \gls{xai} metrics into \texttt{fastSHAP-C} algorithm.
Key \gls{xai} metrics incorporated are:
\begin{enumerate}[noitemsep, nolistsep, label=\large\protect\textcircled{\small\arabic*}]

\item \textit{Confidence Score $\mathcal{CS}$:}  $\mathcal{CS}$ is assessing the reliability and robustness of the explanations generated by \texttt{fastSHAP-C}.
It measures the consistency between the model's predictions and the explanations provided by the model.
A high $\mathcal{CS}$ indicates that the explanations are accurate and align closely with the model's behavior, fostering trust in the model's predictions.
Let \( f(x) \) denote the model's prediction for input \( x \), and \( \phi(x) \) represent the vector of feature importances (Shapley values) provided by \texttt{fastSHAP-C} for input \( x \).
The reconstructed prediction using the explanations is given by
\begin{equation}
    \hat{f}(x) = \sum_{i=1}^d \phi_i(x)
    \label{eq:reconstructed_prediction}
\end{equation}
where \( d \) is the number of features.
$\mathcal{CS}$ is defined as the average absolute difference between the model's actual prediction and the reconstructed prediction
\begin{equation}
    \text{CS} = \frac{1}{n} \sum_{i=1}^n \left| f(x_i) - \hat{f}(x_i) \right|
    \label{eq:confidence_score}
\end{equation}
where \( n \) is the number of samples, \( x_i \) is the \( i \)-th sample, and \( \hat{f}(x_i) \) is the reconstructed prediction for \( x_i \).
A lower  $\mathcal{CS}$ signifies higher reliability of the explanations, as the reconstructed predictions closely match the actual model outputs.

\item \textit{Sensitivity:} It quantifies the degree to which the explanations provided by \texttt{fastSHAP-C} are affected by small changes in the input data or model parameters.
It assesses the robustness of the model by evaluating how much the explanations vary in response to perturbations in the input features.
For input \( x_i \), the sensitivity \( \lambda(x_i) \) is defined as
\begin{equation}
    \lambda(x_i) = \max_{x_j \in B_\epsilon(x_i)} \frac{ \left\| \phi(x_i) - \phi(x_j) \right\|_2 }{ \left\| x_i - x_j \right\|_2 }
    \label{eq:sensitivity}
\end{equation}
where
\begin{itemize}
    \item \( B_\epsilon(x_i) \) is an \( \epsilon \)-neighborhood around \( x_i \), representing inputs within a small perturbation of \( x_i \).
    \item \( \phi(x_i) \) and \( \phi(x_j) \) are the explanations for \( x_i \) and \( x_j \), respectively.
    \item \( \left\| \cdot \right\|_2 \) denotes the Euclidean norm.
\end{itemize}

A lower sensitivity value indicates that the explanations are stable and robust to small input perturbations, which is essential in dynamic O-RAN environments where network conditions can fluctuate.

\item \textit{Log-odds:} Metric evaluates the importance or relevance of each feature in influencing the model's predictions.
It measures the change in the log-odds of the predicted outcome when a particular feature is present or absent, providing insights into the feature's impact on the model's decision-making process.
For a feature \( p \), the log-odds is computed as
    \begin{equation}
            log-odds(p) = -\frac{1}{L}\sum_{i=1}^{L} log\frac{Pr(\hat{y}|x_i^{(p)})}{Pr(\hat{y}|x_i)}\label{logodds}
    \end{equation}
where
\begin{itemize}
    \item \( L \) is the total number of samples.
    \item \( x_i^{(p)} \) is the input \( x_i \) with feature \( p \) removed or altered.
    \item \( \hat{y}_i \) is the model's predicted output for input \( x_i \).
    \item \( \Pr( \hat{y}_i | x_i^{(p)} ) \) and \( \Pr( \hat{y}_i | x_i ) \) are the probabilities of the predicted outcome given the modified and original inputs, respectively.
\end{itemize}
A positive $log-odds$ value indicates that the presence of feature \( p \) increases the likelihood of the predicted outcome, whereas a negative value suggests it decreases the likelihood.
\end{enumerate}

\subsection{Hyper-parameter Tuning}

Hyper-parameter tuning is a critical step in developing effective  learning models, particularly for complex architectures like autoencoders.
A well-structured tuning process ensures that finding the optimal combination of parameters that yields the best model performance.
Robust tuning can significantly improve the performance of the model  leading to more accurate anomaly detection and better generalization.
Furthermore, optimized hyper-parameters help the model converge faster and more efficiently, saving computational resources for resource-constrained O-RAN networks.
It helps prevent overfitting or underfitting by finding hyper-parameters that strike the right balance between model complexity and learning capability.
Given the complexity of our problem, systematic method called GridSearchCV \cite{pedregosa2011scikit} is selected in our tuning process. Grid search automates the search for optimal hyper-parameters by exhaustively testing all combinations in a predefined parameter grid.
The parameters to be tuned and their respective ranges (parameter space) are seen in \Cref{hyperparamxx}. Selected input size of 20, corresponding to the full set of available features in our dataset.

\begin{table}[h]
    \vspace{-.8em}
    \centering
    \scalebox{0.70}{
    \begin{tabular}{@{}lll@{}}
    \toprule
        Hyper-param.
& Search Space & Selected Parameter \\
         \midrule
        Input size  & [2, 4, 8, 16, 20, 32] & 20 \\
        Hidden layers & [1, 2, 3] & 2 \\
        Hidden layer s.& [0.5, 1.5, 2.0, 3.0, 5.0]	 & 3.0 $x$ Input Size\\
        Learning rate & [0.0001, 0.0005, 0.001, 0.005, 0.01] & 0.001\\
        Batch size & [32, 64, 128, 256, 512] & 64 (Mini-batch) \\
        Reconstruction Loss & [MAE, MSE, sMAPE] & MSE \\
        Training steps & [1000, 5000, 10,000, 20,000, 40,000] & 10,000 \\
        Optimizer & [RMSprop, SGD, Adam] & Adam \\
    \bottomrule
    \end{tabular}
 }
    \caption{Hyper-parameter tuning stage of \texttt{SS-DeepCAE}}
    \label{hyperparamxx}
\end{table}

Considering that each feature may carry significant information relevant to anomaly detection in O-RAN traffic, it was crucial to utilize the complete feature set to capture the complex patterns present in high-dimensional data.
GridSearchCV decides 2 hidden layers in the encoder and decoder parts.
This choice balances model complexity and representational capacity.
Mathematically, deeper networks can approximate more complex functions due to their ability to represent higher-order interactions among features \cite{hornik1989multilayer}.
However, increasing the number of layers also increases the risk of overfitting and computational cost.

In the context of O-RAN, where real-time processing is essential, a model with 2 hidden layers provides sufficient depth to learn meaningful representations while maintaining computational efficiency.
Hidden layer size selected to be 3 times the input size, resulting in 60 neurons for the first hidden layer.
This scaling ensures that the network has enough capacity to capture the intricate structures in the data without introducing unnecessary parameters.
The subsequent layers reduce the dimensionality, forming a bottleneck that forces the model to learn compressed representations.
A learning rate of 0.001 was chosen, which is a common default value for the Adam optimizer.
This rate offers a good balance between convergence speed and the stability of the optimization process.
A too high learning rate might cause the optimizer to overshoot minima, while a too low rate could result in slow convergence or getting stuck in local minima.
Mini-batch size selected 64.
Mini-batch gradient descent helps in smoothing the optimization landscape and accelerates convergence compared to stochastic gradient descent \cite{bottou2010large-scale}.
A batch size of 64 provides a good trade-off between computational efficiency and the robustness of parameter updates.

Mean Squared Error (MSE) was selected as the reconstruction loss function.
MSE penalizes larger errors more than smaller ones due to the squaring term, making it suitable for capturing significant deviations in reconstruction, which is critical in anomaly detection.
Adam optimizer was chosen for its adaptive learning rate capabilities and generally good performance across various tasks.
It combines the advantages of both AdaGrad and RMSprop, adjusting the learning rate for each parameter dynamically, which is beneficial for training deep neural networks.
Rectified Linear Units (ReLU) were used as the activation function in all layers except the output layer.
 ReLU is simple to compute, leading to faster training times.
Unlike sigmoid or tanh activations, ReLU does not saturate in the positive region, helping to mitigate the vanishing gradient problem in deeper networks \cite{glorot2011deep}.
Also, ReLU induces sparsity in activations, as negative inputs are mapped to zero.
This sparsity can enhance feature learning and reduce overfitting.
The total number of trainable parameters is 8,180, which is relatively low given the model's capacity.
This low complexity is advantageous for deployment in \gls{oran} systems, where computational resources may be limited, and low latency is essential.

\begin{table}
    \centering
    \scalebox{0.75}{

        \begin{tabular}{lll}
        \toprule
        \textbf{\texttt{SS-DeepCAE} Layers} & \textbf{Output Shape}   & \textbf{Parameters}    \\
        \midrule
        input\_1 (InputLayer)               & (None, 20) & 0 \\
        dense\_1 (encoded)                  & (None, 64) & 1344  \\
        dense\_2 (encoded)                  & (None, 32) & 2080  \\
        dense\_3 (bottleneck)               & (None, 16) & 528  \\
        dense\_4 (bottleneck)               & (None, 16) & 272  \\
        dense\_5 (decoded)                  & (None, 32) & 544  \\
        dense\_6 (decoded)                  & (None, 64) & 2112  \\
        dense\_7 (Dense)                    & (None, 20) & 1300  \\
        \midrule
        \textbf{Total Params:}              & 24,452     &  \\
        \textbf{Trainable Params:}          & 8,180      &  \\
        \textbf{Non\-trainable Params:}     & 0          &  \\
        \textbf{Optimizer Params:}          & 16,362     &  \\
        \bottomrule
     \end{tabular}
}
    \caption{\texttt{SS-DeepCAE} model summary with parameters}
    \label{deepCAE}
    \vspace{-.8em}
\end{table}

The \gls{oran} environment presents unique challenges due to its disaggregated architecture and the heterogeneity of components.
High-dimensional data streams require models that can effectively capture complex patterns without incurring significant computational overhead.
Our \texttt{SS-DeepCAE} model in \Cref{deepCAE} addresses these challenges by; 2 encoder and 2 decoder layers provide sufficient depth to model non-linear relationships in the data while keeping the model lightweight.
 The bottleneck layer compresses data to a lower-dimensional space, which is critical for identifying anomalies as deviations from learned normal patterns.
Contractive Loss enhances the model's robustness to noise and variations in the input, which is common in real-world network traffic data.

\subsection{Baseline Models}
\subsubsection{Baseline Models for our \texttt{SS-DeepCAE} Model}

All baseline models whose performance we compared with \texttt{SS-DeepCAE} model  implemented as semi-supervised.
We consider the following baselines:
\begin{enumerate}
    \item Vanilla Autoencoder: Vanilla autoencoder can be described in it is most basic form as a neural network comprising three layers, which includes a single hidden layer.
The input and output of this network are identical, and the objective is to learn the process of reconstructing the input.
This is typically achieved through the utilization of the Adam optimizer in conjunction with the mean squared error loss function.
    \item LSTM-Autoencoder \cite{borghesi2019anomaly}: This architecture consists of three layers; an LSTM encoder featuring 256 units with a dropout rate of 20\%, an LSTM decoder incorporating 512 units also with a 20\% dropout rate, and a dense output layer that matches the number of units to the length of the small time series.

    \item SLA-VAE \cite{huang2022vae}: Model defines anomalies
    based on their feature extraction module, then introduces semi-supervised
    VAE to identify anomalies in multivariate time series.
Authors adopt the reconstruction error to detect anomalies.
Their module consists of two steps.
First, they compute the reconstructed output with semi-supervised VAE, and then
    calculate the reconstruction error for each observation.
    \item DeepAE \cite{basaran2023deep}: It is our first model that helps us understand the problems brought by DeepAE and motivated us to develop new scalable, less complex model which we clarified in \Cref{sec:intro}.
It is a high-power deep autoencoder model with a structure consisting of shallow layers to be symmetrical in the encoder and decoder parts.
It has $30,491$ trainable, $806$ non-trainable, $60,984$ optimizer and $92.281$ total parameters.
Compared to proposed \texttt{SS-DeepCAE}, it has almost $4$ times more parameters.
\end{enumerate}

\subsubsection{Baseline Models for our XAI Method \texttt{fastSHAP-C}} \label{xaibaselines}
We consider the following baselines for our \texttt{fastSHAP-C} XAI algorithm:
\begin{enumerate}
    \item kernelSHAP \cite{lundberg2017unified}: This approach involves analyzing the predictions made by machine learning models through the lens of Shapley values, which originate from cooperative game theory.
The primary objective is to equitably assign the contribution of each feature to the overall prediction of the machine learning model.
    \item fastSHAP \cite{jethani2022fastshap}: Algorithm serves as an effective approximation technique for calculating Shapley values.
Its primary objective is to expedite the computation of Shapley values, which can be resource-intensive when dealing with intricate models and extensive datasets.
fastSHAP utilizes a differentiable surrogate model that is trained to estimate Shapley values, thereby considerably decreasing the time required for computation.
\end{enumerate}

%

\section{Results and Discussion}
\label{sec:results}

In this section, we analyze \texttt{SS-DeepCAE} model's performance.
Then continue by interpreting the SHAP values to gain insights into the behavior of the \texttt{SS-DeepCAE} model, which functions as an \gls{adxapp}.
This analysis helps reveal how specific features contribute to the model's anomaly detection decisions, enhancing our understanding of it is inner workings.
Following this interpretability assessment, we conduct a comprehensive performance benchmarking of our novel \texttt{fastSHAP-C} implementation within the XAInomaly framework.
We compare the results against kernelSHAP and fastSHAP, which are commonly utilized in recent research, to evaluate \texttt{fastSHAP-C}'s effectiveness and improvements in both speed and resource efficiency.

\subsection{Performance Results of \texttt{SS-DeepCAE} Model}

\subsubsection{Training and Validation Performance}

\begin{figure}
    \centering
    \subfloat[Training/Validation Loss.]
    {\includegraphics[width=0.8\columnwidth]{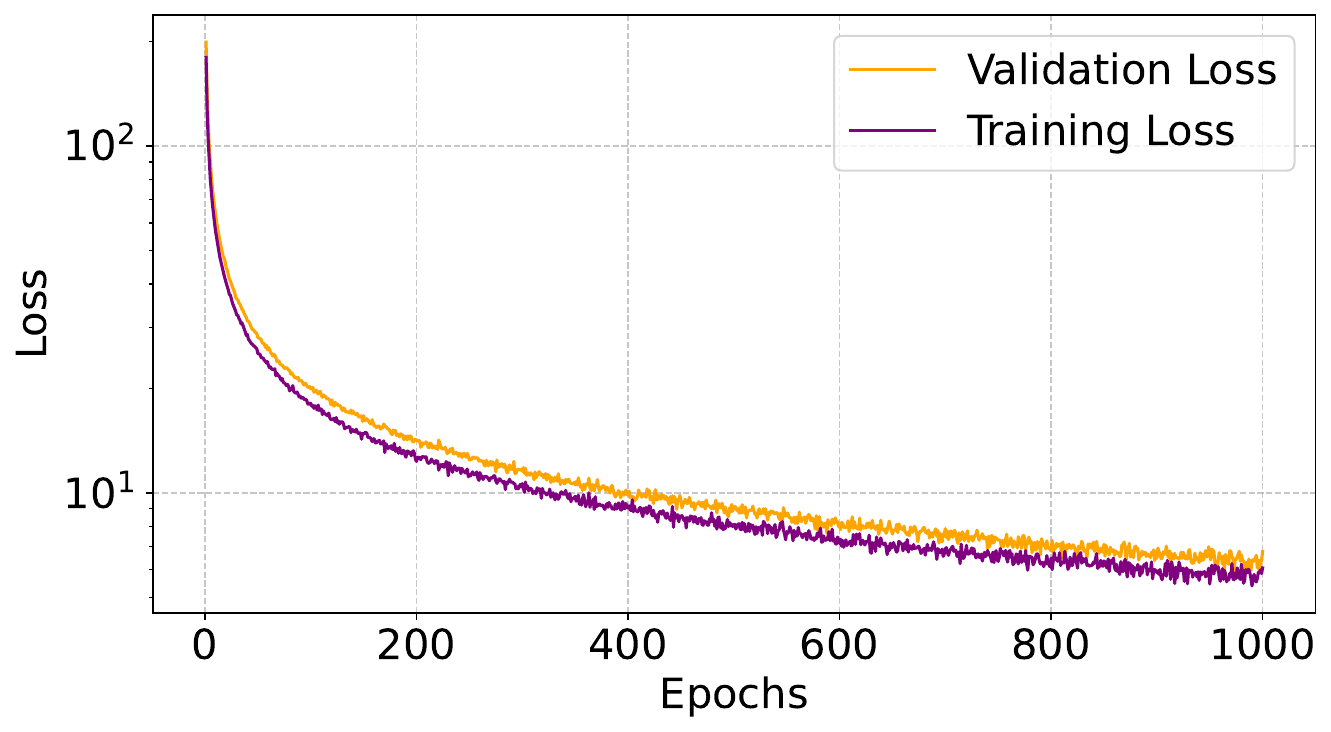}\label{trainvalloss}\label{fig_losses_a}}
    \hfill
    \subfloat[Training/Validation Accuracy.]
    {\includegraphics[width=0.8\columnwidth]{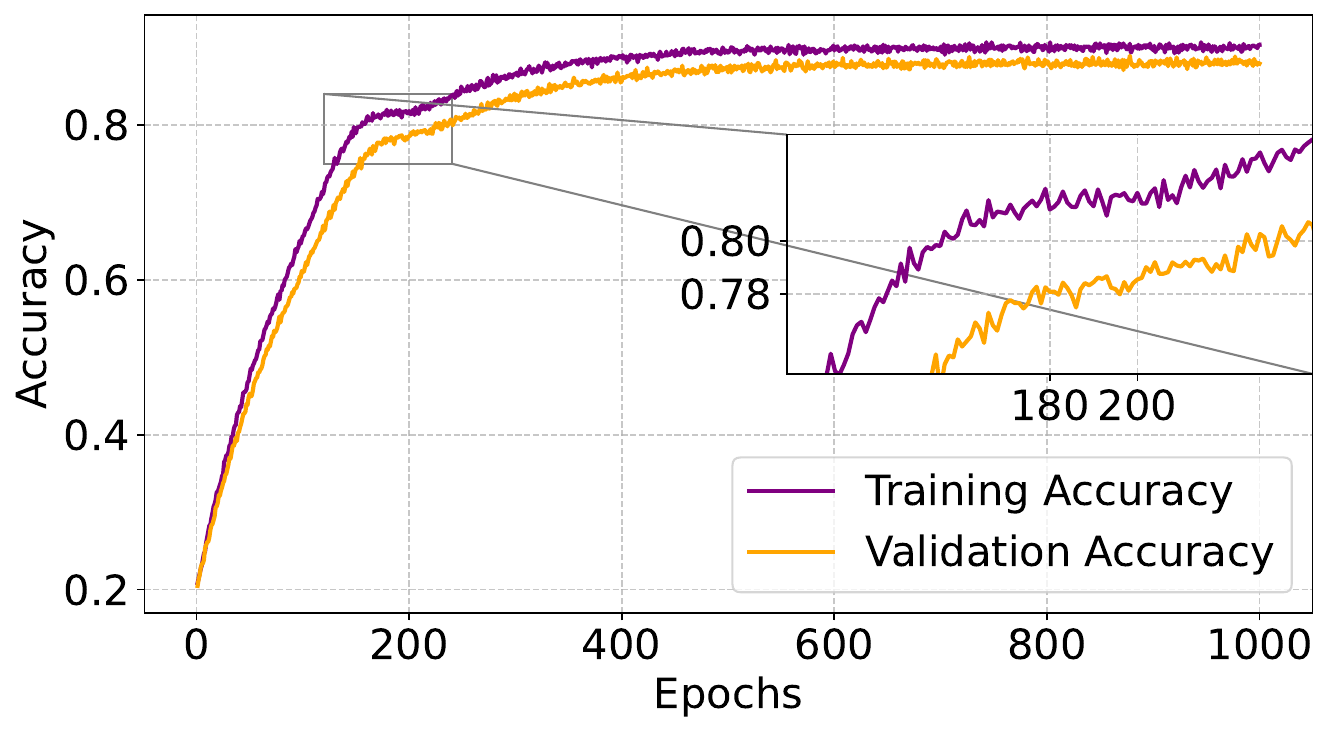}\label{accuracy}\label{fig_losses_b}}
    \caption{Training and validation loss-accuracy curves over epochs}
    \label{fig_losses}
    \vspace{-.8em}
\end{figure}

The performance of the \texttt{SS-DeepCAE model} is visualized through the loss and accuracy curves over the training epochs as seen in \Cref{fig_losses}, providing insights into the model's learning behavior and generalization capabilities.
Both the training and validation loss curves, plotted on a logarithmic scale, demonstrate a steep decline during the initial epochs, indicative of the model effectively capturing the foundational patterns in the data.
This behavior is mirrored in the accuracy curves, where both training and validation accuracy exhibit a rapid increase in the early stages, highlighting the model's efficiency in learning critical features for anomaly detection.
As training progresses, the slopes of the loss curves begin to flatten after approximately 200 epochs, signaling a transition to fine-tuning the model's representations.
By around 600 epochs, the loss curves stabilize, indicating convergence.
Notably, the training and validation loss curves remain closely aligned throughout the training process, which suggests that the model is not overfitting.

This alignment is particularly important in real-world O-RAN AI/ML models deployment, where generalization to unseen data is critical.
The accuracy curves provide further evidence of the model's strong performance.
Both training and validation accuracy improve steadily, with the training accuracy eventually reaching 90\% and validation accuracy stabilizing at approximately 82\%.
The gap between these curves remains small, demonstrating good generalization without overfitting.
The zoomed section of the accuracy curve, focusing on epochs 180 to 200, reveals a critical learning phase where validation accuracy rapidly improves from approximately 70\% to 80\%.
This sharp increase suggests that the model transitions to capturing more complex or higher-order patterns in the data during this phase.

\subsubsection{Analysis of Reconstruction Loss and Precision with Varying Layer Sizes}

\begin{figure}
    \centering
    \includegraphics[width=0.95\columnwidth]{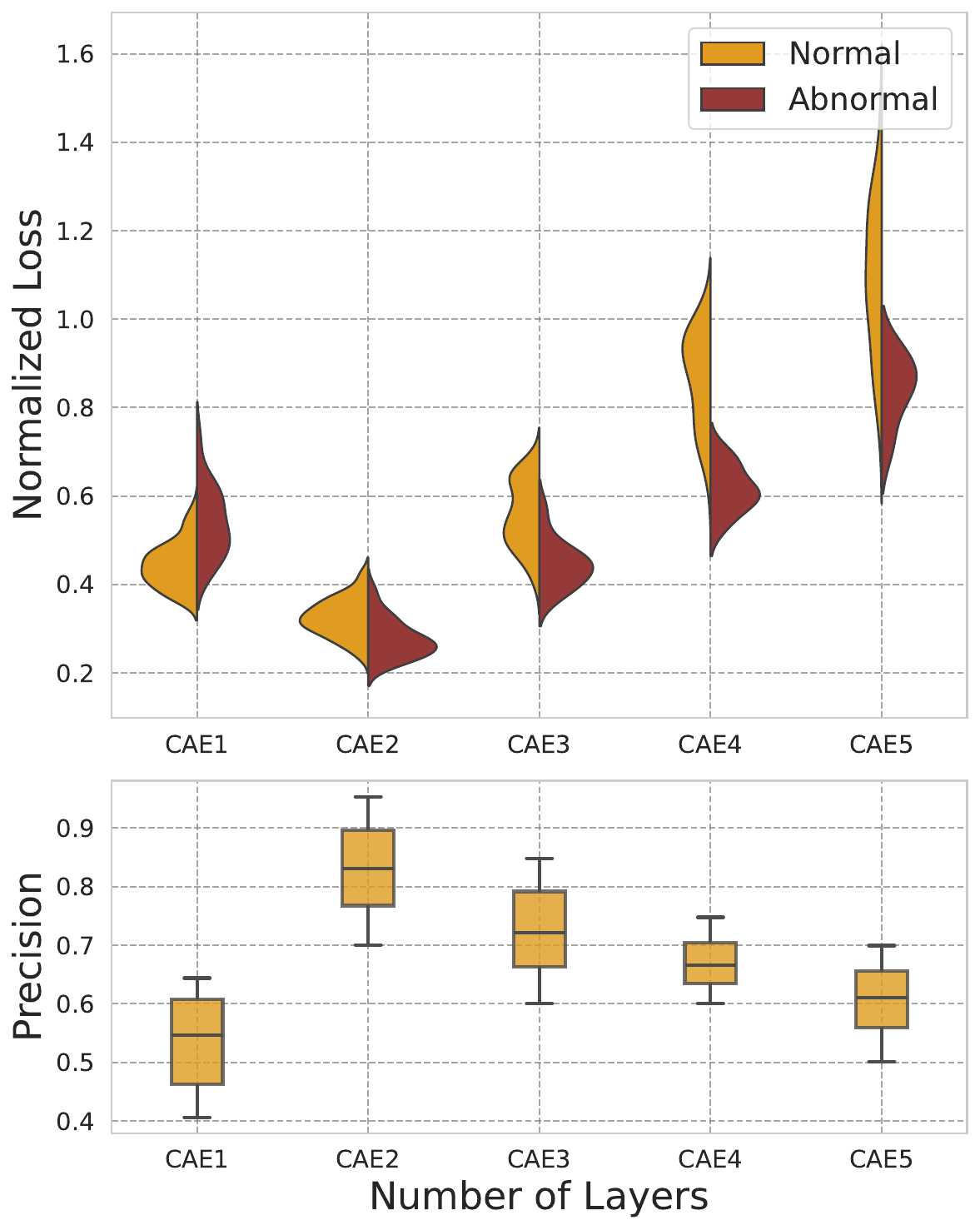}
    \caption{Reconstruction error distribution for \texttt{SS-DeepCAE} with different layer size (Upper Graph) Precision distribution of \texttt{SS-DeepCAE} with different layer size (Lower Graph)}
    \label{layers}
    \vspace{-.8em}
\end{figure}

In \Cref{layers}, the provided graphs illustrate the performance of the \texttt{SS-DeepCAE} model across different layer configurations (\texttt{CAE1} to \texttt{CAE5}) in terms of reconstruction loss and precision.
The results clearly indicate that \texttt{CAE2}, which represents a model with two layers in both the encoder and decoder, delivers the best trade-off between reconstruction error and precision.

The reconstruction loss, depicted in the upper graph, compares the model's ability to differentiate between normal (yellow) and abnormal (red) samples across various configurations.
\texttt{CAE2} exhibits the narrowest distribution of reconstruction errors for both normal and abnormal samples, indicating it is superior ability to capture the underlying structure of normal network behavior while effectively separating abnormal instances.
Smaller spread of the reconstruction loss for \texttt{CAE2} suggests that it provides a robust and consistent representation of the input data, minimizing overfitting while maintaining sensitivity to anomalies.
As the number of layers increases (\texttt{CAE3}, \texttt{CAE4}, \texttt{CAE5}), the reconstruction error for both normal and abnormal samples grows, and the distributions widen.
This behavior is indicative of increased model complexity, which introduces noise and reduces the model's ability to generalize effectively.
For \texttt{CAE5}, the reconstruction loss overlaps significantly between normal and abnormal samples, showing the model's reduced ability to distinguish anomalies.
This suggests that the added complexity fails to improve the latent representations and instead hampers the model's performance.

The precision distribution, shown in the lower graph, evaluates the model's ability to correctly identify anomalies as the layer size varies.
\texttt{CAE2} demonstrates the highest median precision, along with the tightest interquartile range (IQR).
This indicates that the two-layer architecture achieves the optimal balance between model capacity and generalization, making it the most effective configuration for anomaly detection.
As the number of layers increases beyond two, the precision drops significantly.
The wider IQRs in these configurations indicate variability in model performance, which may be attributed to overfitting or the inability to effectively optimize the additional parameters introduced by deeper architectures.
Particularly in \texttt{CAE5}, the median precision is markedly lower, and the distribution indicates poor anomaly detection capability.
This reinforces that additional layers do not necessarily translate to better performance and may instead lead to diminished returns.
While \texttt{CAE1} avoids overfitting by keeping the architecture simple, it is lower precision compared to \texttt{CAE2} highlights the trade-off between model complexity and expressive power.
A single-layer configuration lacks sufficient depth to capture the complex relationships in high-dimensional \gls{oran} data, leading to missed anomalies and reduced precision.

\subsubsection{Performance Comparison with Baseline Models}

 \begin{figure*}
        \subfloat[Accuracy Results]{%
            \includegraphics[width=.48\linewidth]{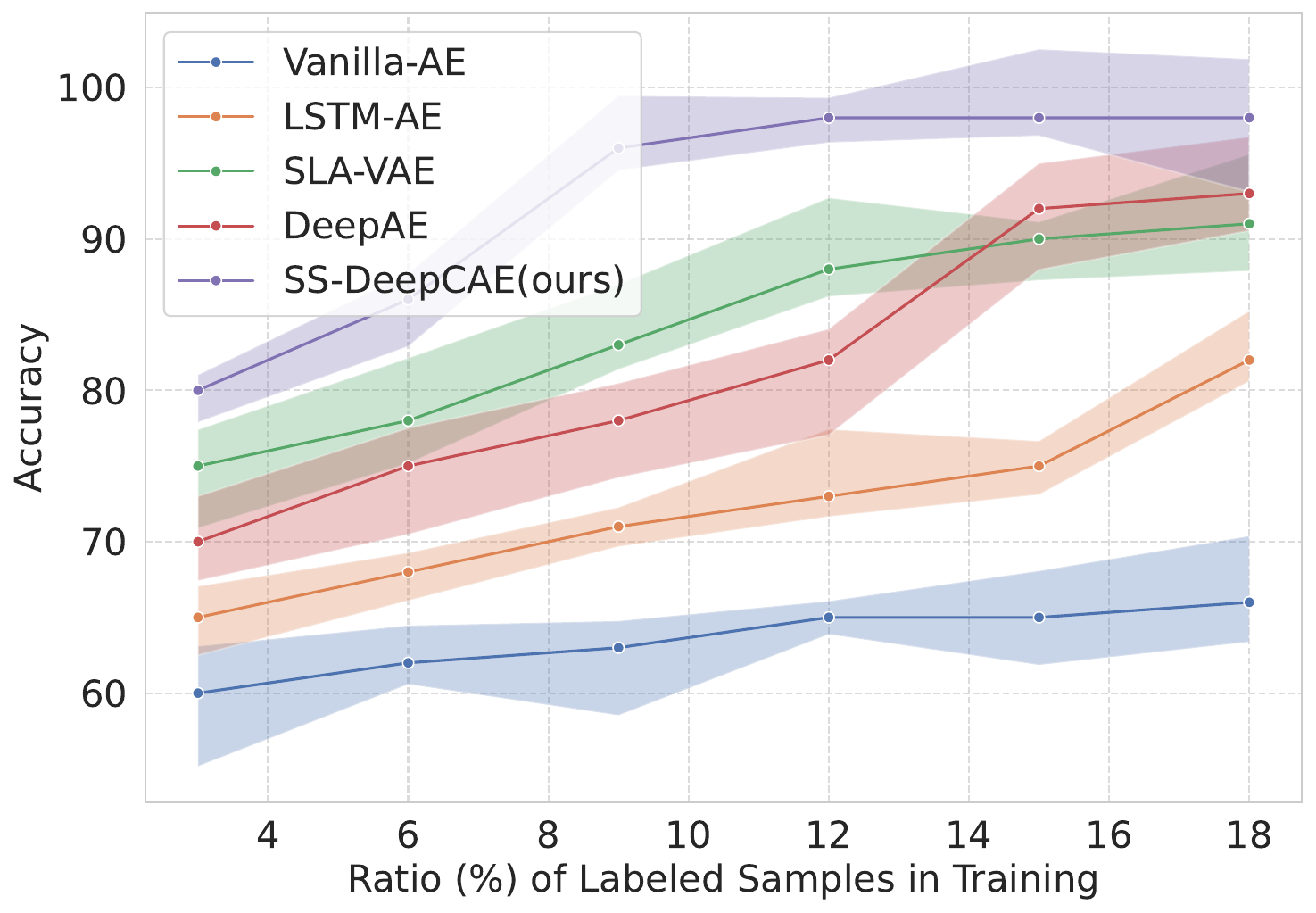}%
            \label{subfig:a}%
        }\hfill
        \subfloat[Precision Results]{%
            \includegraphics[width=.48\linewidth]{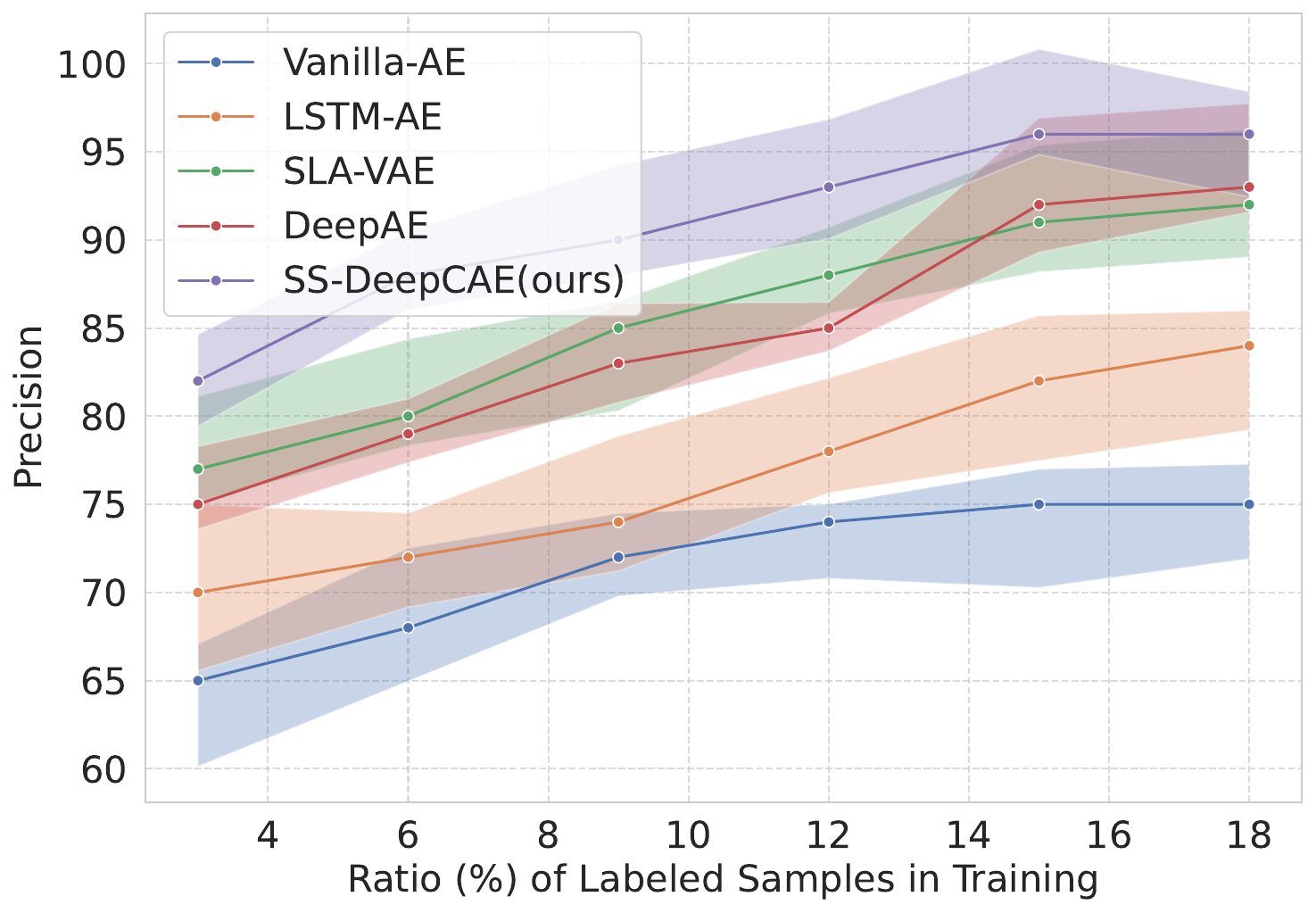}%
            \label{subfig:b}%
        }\\
        \subfloat[F1-Score Results]{%
            \includegraphics[width=.48\linewidth]{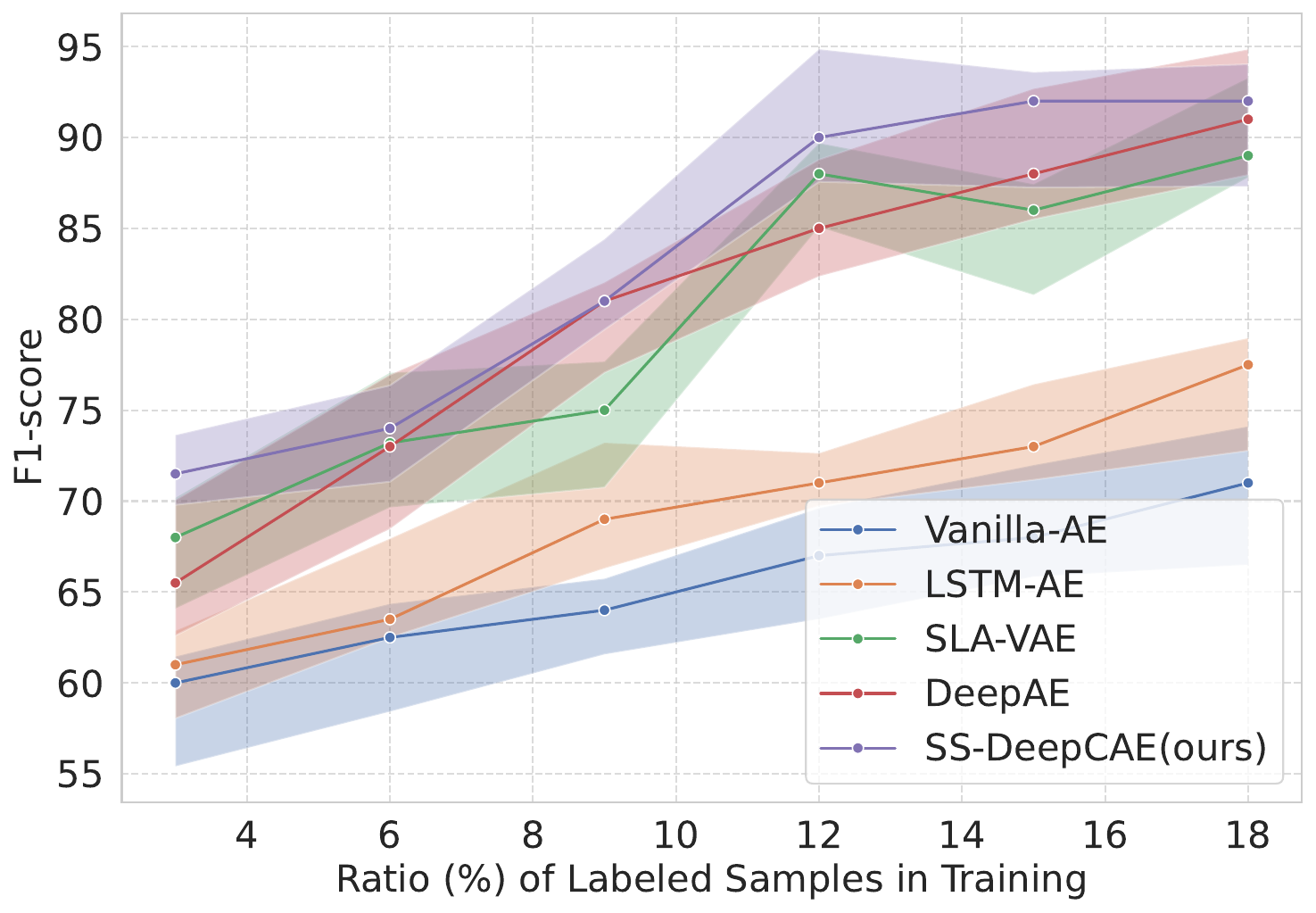}%
            \label{subfig:c}%
        }\hfill
        \subfloat[AUC Results]{%
            \includegraphics[width=.48\linewidth]{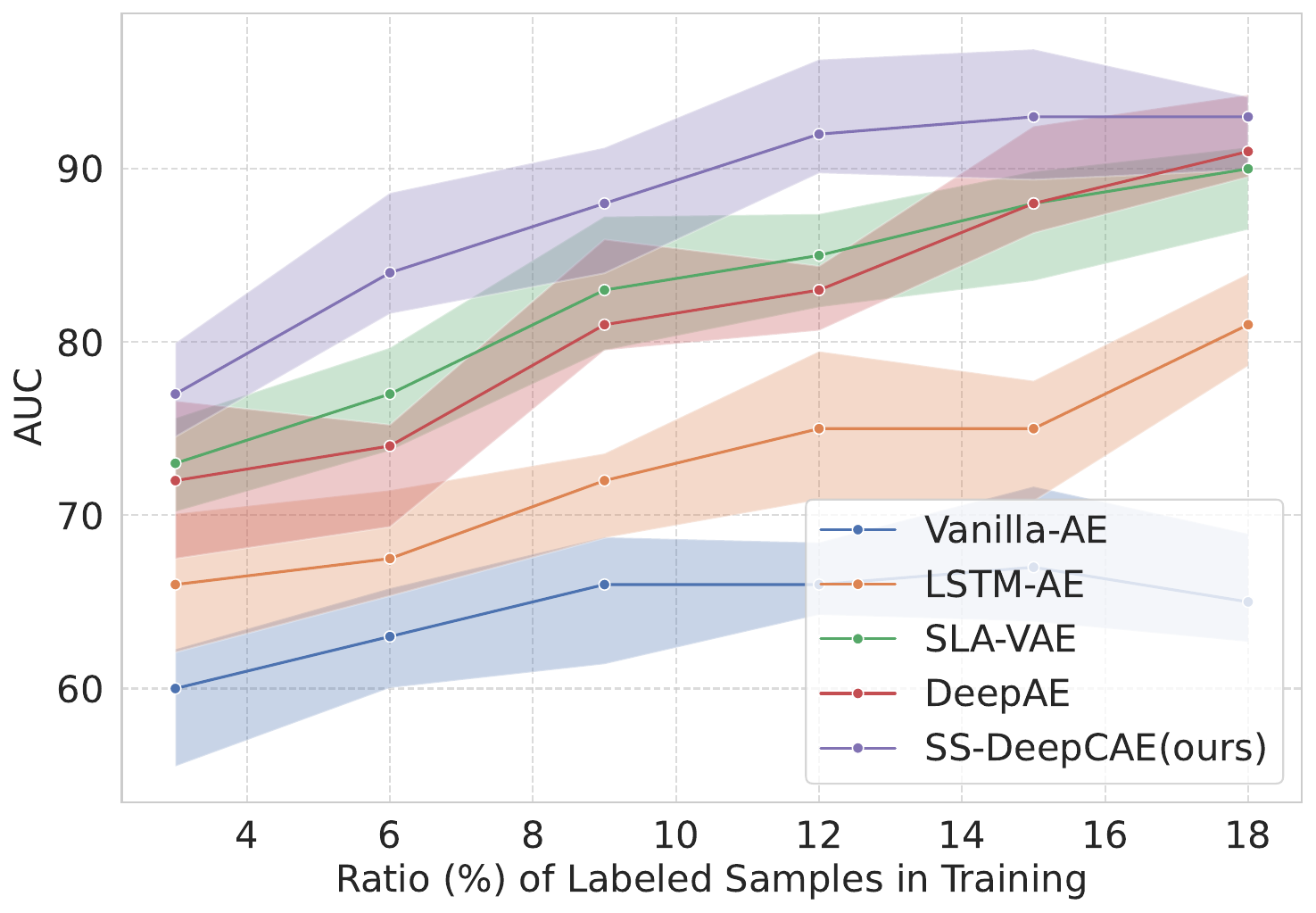}%
            \label{subfig:d}%
        }
        \caption{Accuracy, Precision, F1-score, and AUC variation curves of models with different ratio labeled samples}
        \label{performanceresults}
\end{figure*}

\begin{table*}
\centering
   \scalebox{0.90}{
    \renewcommand\thetable{IV}
        \begin{tabular}{*{12}{l}}
    \toprule
    &   \multicolumn{6}{c}{\# of Labeled Samples from Dataset}
 \\
    \cmidrule(r) {2-7}

Models
    &   100   &   300    &   500 &   800   &   1000    &   \# All
  \\
    \midrule
Vanilla Autoencoder & $43.46_{\pm (2.1)} $  &  $46.67_{\pm (2.2)} $         & $48.14_{\pm (0.8)} $   &  $51.27_{\pm (2.3)} $   &   $58.73_{\pm (0.9)}$    &   $65.13_{\pm (1.6)} $
     \\
LSTM-Autoencoder \cite{borghesi2019anomaly}  & $59.82_{\pm (1.3)} $  &  $62.22_{\pm (1.8)} $         & $65.71_{\pm (0.9)} $   &  $68.17_{\pm (0.5)} $   &   $71.33_{\pm (2.1)}$    &   $76.63_{\pm (0.9)} $
  \\
SLA-VAE \cite{huang2022vae} & $64.56_{\pm (1.9)} $  &  $68.57_{\pm (1.2)} $         & $71.32_{\pm (0.7)} $   &  $75.19_{\pm (2.1)} $   &   $81.13_{\pm (1.1)}$    &   $85.43_{\pm (1.3)} $
     \\
DeepAE \cite{basaran2023deep}  & $71.72_{\pm (1.5)} $  &  $74.48_{\pm (1.9)} $         & $75.72_{\pm (2.1)} $   &  $76.39_{\pm (0.7)} $   &   $83.33_{\pm (1.8)}$    &   $88.66_{\pm (1.8)} $     \\
\texttt{SS-DeepCAE} (ours) & $80.17_{\pm (0.6)} $  &  $82.67_{\pm (1.2)} $         & $85.12_{\pm (1.1)} $   &  $86.39_{\pm (1.7)} $   &   $91.33_{\pm (1.4)}$    &   $91.17_{\pm (0.6)} $   \\
    \bottomrule
     \end{tabular}
   }
    \caption{Average UAR with standard deviation on test set (different amount of labeled samples in training)}
    \label{averageuar}
    \vspace{-.8em}
\end{table*}

Provided results evaluates \texttt{SS-DeepCAE} against various baseline architectures, cross four key metrics as can be seen in \Cref{performanceresults}. These metrics are analyzed with respect to the increasing percentage of labeled samples in training.
The results demonstrate that the proposed model consistently outperforms the baselines, particularly in scenarios with limited labeled data, making it a highly effective solution for semi-supervised anomaly detection in \gls{oran} environments.

SS-DeepCAE achieves the highest accuracy, surpassing 95\% with just 10-12\% labeled samples.
It is accuracy consistently outpaces the baselines, particularly in the low-labeled-data regime (4–10\% labeled samples), where it maintains a significant margin over the others.
Superior accuracy is attributed to the contractive penalty in our \texttt{SS-DeepCAE}'s loss function design, which enhances the model's ability to learn robust and smooth feature representations that generalize well to unseen data.
This is especially advantageous in high-dimensional, noisy and lack of labeled data.

\textit{Our model significantly outperforming the baselines at higher labeled ratios.} Even with as little as 6\% labeled data, it maintains precision above 90\%.
Precision measures the fraction of correctly identified anomalies among all instances flagged as anomalous.
The \texttt{SS-DeepCAE}'s ability to reduce false positives is enhanced by its semi-supervised learning approach, which effectively separates normal and abnormal traffic patterns in the latent space.
DeepAE performs well at higher labeled data ratios but struggles in the low-labeled regime, indicating its tendency to overfit to the available labeled data.
Vanilla-AE and SLA-VAE lag behind due to their inability to handle high-dimensional data efficiently without extensive supervision.

F1-score captures the trade-off between precision and recall.
\texttt{SS-DeepCAE}'s contractive loss enables the model to maintain high recall by learning representations that are robust to minor variations in the input, ensuring that true anomalies are not overlooked.
High AUC underscores the effectiveness of the bottleneck layer in the \texttt{SS-DeepCAE} architecture, which compresses high-dimensional input data into a meaningful latent representation, enabling the model to consistently rank anomalies higher than normal samples.

In \Cref{averageuar},  we present UAR results for six labeled sample configurations, which capture a range starting from a small number of labeled samples (100) up to the total set, illustrating how \texttt{SS-DeepCAE} evolves from a data-scarce scenario to one where abundant labeled data is available. We focus on UAR at these discrete points because it robustly highlights performance in imbalanced conditions, especially relevant for anomaly detection, where the minority class (anomalies) can be overwhelmed by normal data. By showcasing UAR at successively larger subsets of labeled data, we illustrate our model's trajectory in learning critical features of anomalous samples under varying degrees of label availability. Meanwhile, Accuracy, Precision, F1-score, and AUC are also reported at lower labeled ratios (down to 4–10\%) to underscore the consistency of our approach across different metrics and further validate the effectiveness of \texttt{SS-DeepCAE}. Taken together, these evaluations provide a comprehensive view of how the model scales and performs as labeled sample sizes incrementally increase, rather than focusing only on full labeled samples. UAR results emphasize the superiority of our model, particularly in scenarios with limited labeled data, and its ability to converge to high performance levels without requiring substantial increases in labeled data.

\texttt{SS-DeepCAE} consistently outperforms all baseline models across all labeled sample sizes.
With only 100 labeled samples, \texttt{SS-DeepCAE} achieves a UAR of $80.17 \pm 0.6$, significantly higher than the next best-performing model, DeepAE $(71.72 \pm 1.5)$.
This trend persists as the number of labeled samples increases, culminating in $91.17 \pm 0.6$ when using the full dataset, outperforming DeepAE $(88.66 \pm 1.8)$ and SLA-VAE $(85.43 \pm 1.3)$.
Performance of \texttt{SS-DeepCAE} begins to stabilize after 800 labeled samples, with a marginal increase in UAR beyond this point.
For instance,  UAR improves from $86.39 \pm 1.4$ (800 labeled samples) to $91.17 \pm 0.6$ (full dataset).

This saturation suggests that the \texttt{SS-DeepCAE} effectively utilizes both labeled and unlabeled data, achieving robust representations that do not heavily rely on additional labeled samples.
Beyond 800 labeled samples, the model's UAR begins to plateau, indicating that the additional labeled data provides diminishing returns.
This suggests that the model's reliance on labeled data is significantly reduced, making it ideal for scenarios where labeling is expensive or infeasible.
The ability of \texttt{SS-DeepCAE} to achieve high UAR with minimal labeled data makes it highly scalable for O-RAN environments, where labeled anomalies are rare and costly to obtain.
This efficiency ensures that the model can be deployed quickly without the need for extensive labeling efforts.

\subsubsection{Model Complexity Analysis for Resource-Constrained Environments}
While our proposed \texttt{SS-DeepCAE} model significantly reduces the number of trainable parameters compared to standard DeeAEs, it is still crucial to assess its computational footprint for real-world \gls{oran} deployments. A detailed breakdown of the total parameters and memory usage is given in \cref{deepCAE,tab:performancexxx}, illustrating that the encoder-decoder structure with contractive regularization strikes a balance between model capacity and efficiency. In resource-constrained environments (e.g., edge nodes or small-footprint \gls{nearrt} deployments), three aspects are particularly relevant:

\textit{Parameters and Memory Usage:} Fewer network layers and careful bottleneck design decrease the total number of parameters, thereby lowering memory requirements. Our architecture keeps the parameters at a level that remains feasible for on-device or near-edge processing without necessitating large GPU clusters.

\textit{Inference Latency:} The contractive penalty induces smoother and more robust representations, helping the model converge faster at inference time. However, deeper models, if deployed, will incur higher latency. Depending on the desired latency budget (e.g., sub-second decisions in \gls{oran} anomaly detection), there may be a need to prune layers or reduce dimensionality.

\textit{Model Convergence Time:} In some O-RAN scenarios, particularly in micro-data centers or edge servers, limited energy availability constrains continuous training. By using contractive regularization, our model tends to converge in fewer epochs and can be partially fine-tuned or updated incrementally, mitigating energy consumption overhead.

For future complexity trade-offs, employing adaptive network architectures, where deeper layers are skipped if the detection confidence is sufficiently high, offers a route to adjust computational complexity based on real-time demands. Tailoring \texttt{SS-DeepCAE} through pruning weights or entire neurons associated with lower-priority sub-tasks can significantly reduce overhead while retaining high accuracy.

\subsection{Performance Results of \texttt{fastSHAP-C}}
In this section, we particularly examine the explainability and interpretability capacity of proposed \texttt{fastSHAP-C} algorithm and compared its performance with baseline models specified in \Cref{xaibaselines}.
We also ran our \texttt{fastSHAP-C} method on our new anomaly detection model \texttt{SS-DeepCAE} and observed the results.

\subsubsection{SHAP Values and Feature Contributions}

\begin{figure}
    \centering
    \includegraphics[width=\columnwidth]{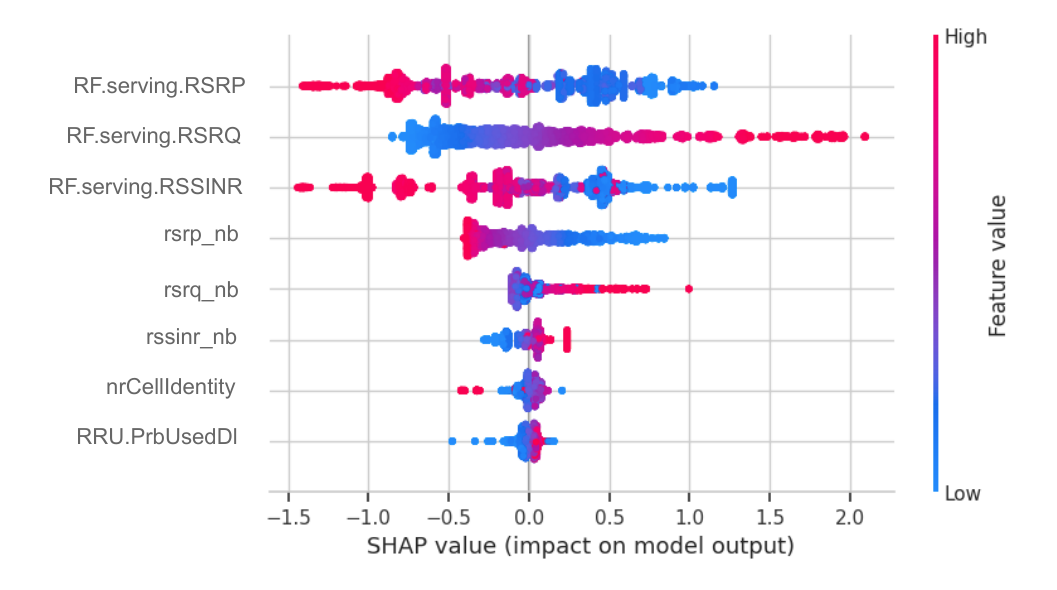}
    \caption{Explaining feature contributions with SHAP values}
    \label{shapv}
    \vspace{-.8em}
\end{figure}

The process of calculating SHAP values is typically NP-hard, which can result in high convergence times and increased complexity in obtaining solutions.
To address this challenge, we implemented kernelSHAP as our baseline model to compute SHAP values and gain insights into how each feature contributes to the model's predictions.
Starting with results in \cref{shapv}, we present a SHAP summary plot that depicts the contribution of each input feature to our \texttt{SS-DeepCAE} anomaly detection model's output. On the horizontal axis, the “SHAP value” measures how much each feature, for a given sample, shifts the model's prediction toward anomalous (positive SHAP) or toward normal (negative SHAP). Each point represents a specific instance in the dataset, color-coded from blue (lower feature values) to red (higher feature values). Features at the top of  \cref{shapv} are generally those that, on average, have the greatest overall impact on the model's decisions; features appearing toward the bottom have less global influence or tend to have specialized significance for particular types of anomalies. The most influential metrics in this O-RAN setting emerge as “RF.serving.RSRP,” “RF.serving.RSRQ,” and “RF.serving.RSSINR.” These correspond to critical \gls{rf} measurements from the serving cell.  Observing the SHAP values on \cref{shapv}, data points with lower RSRP (blue coloring) tend to push the model toward labeling the sample as an anomaly (larger positive SHAP on the horizontal axis). This aligns with real-world domain knowledge: coverage holes or sub-optimal power levels often manifest as performance degradations that trigger anomaly alerts. 

Likewise, RF.serving.RSRQ offers an integrated measure of signal quality by considering both signal strength and interference. In an O-RAN context, when RSRQ is too low or fluctuates drastically, the network's capacity for maintaining stable connections diminishes. As shown in \cref{shapv}, instances with poor RSRQ values also exhibit prominent positive SHAP attributions, confirming that degraded signal quality is a major contributor to anomaly classification. Moreover, RF.serving.RSSINR indicates how much stronger the useful signal is compared to interference plus background noise. Markedly low RSSINR translates to heightened interference scenarios -- an important element in diagnosing anomalies tied to poor spectral efficiency or environmental blockages. When RSSINR is low (indicated by blue or purple in \cref{shapv}), the associated SHAP value tends to push the outcome toward anomalous. Beyond the serving cell metrics, “rsrp\_nb,” “rsrq\_nb,” and “rssinr\_nb” represent analogous RF measurements (RSRP, RSRQ, RSSINR) captured from neighboring cells. These features aid in understanding whether a user equipment (UE) in distress might connect better to nearby cells or if poor coverage is widespread.

In \cref{shapv}, outlier points colored red (high neighbor-RSRP or neighbor-RSSINR) sometimes have negative SHAP values, indicating that strong neighboring signals can reduce anomaly likelihood, even if the serving cell metrics are suboptimal, because the network can potentially initiate a handover. Conversely, low values for these neighbor-cell metrics reinforce an anomaly label, as the UE lacks alternatives to recover service quality. By aligning model explanations with known domain-specific mechanics, SHAP values offers a transparent window into how each feature drives the model's anomaly decisions. Such interpretability is crucial for O-RAN operators who not only require real-time alerts but also actionable insights into why those alerts happen. In practice, these SHAP attributions can inform xApps when adjusting handover thresholds, recalibrating power control, or orchestrating beamforming to counteract the underlying conditions that lead to anomalies.

\begin{table}
    \centering
    \scalebox{0.85}{
        \begin{tabular}{@{}llll@{}}
        \toprule
        \textbf{Features} &    &     &  \\
        \midrule
        High Reconstruction   Error    & \hfil $X_1$   \\
        Explanatory                   & \hfil $X_2, X_3$\\
        Explaining Anomaly                  & \hfil $X_1,X_2,X_3$ \\
        \bottomrule
        \end{tabular}
    }
    \caption{Explainability with model features}
    \label{featurecont}
    \vspace{-.8em}
\end{table}

Results shown in \cref{featurecont} reveal that RF.serving.RSRPv ($X_1$) consistently emerges as the most influential feature determining whether UE behavior is flagged as anomalous. From an explainability standpoint, the high Shapley values associated with $X_1$ indicate that even marginal fluctuations in the Reference Signal Received Power (RSRP) can significantly alter the model's anomaly score. In other words, when $X_1$ shifts from moderate to low values, the model's confidence that the UE is experiencing an anomalous condition increases sharply. This aligns with domain knowledge: as RSRP falls, the UE's link quality deteriorates, potentially leading to dropped connections and handover failures. Similarly, RF.serving.RSRQ ($X_2$) and RF.serving.RSSINR ($X_3$) exhibit notable Shapley contributions, indicating that both metrics frequently influence the final anomaly classification. Specifically, $X_2$ encapsulates Reference Signal Received Quality, which ties signal strength to prevailing interference; low RSRQ therefore suggests an environment of pronounced interference or noise. High feature attributions for $X_2$ imply that the model has learned to associate elevated interference with abnormal network states -- an interpretation that is critical for proactive troubleshooting of congested or interference-prone cells.

In parallel, $X_3$ tracks the Received Signal Strength Indicator to Noise Ratio (RSSINR). Large Shapley values for $X_3$ generally appear when the RSSINR dips below a certain threshold, reinforcing the notion that poor signal-to-noise conditions are a key driver of anomalous UE performance. This effect can be especially acute if $X_1$ is already borderline, meaning that coverage is insufficient to maintain reliable throughput in the face of rising interference. Under these circumstances, the model's explainability outputs highlight that both a weak signal ($X_1$) and high interference ($X_2$, $X_3$) jointly account for triggering anomalies. The combination of $X_1$, $X_2$, and $X_3$ points to a situation in which a single metric alone may not provide a complete explanation for anomalous behavior. Instead, the model relies on how these RF indicators interact.
If one of them degrades notably, the Shapley values for the remaining features can become more or less pronounced, depending on the network context (e.g., mobility events, cell-edge conditions).

Such insight is vital for network administrators who wish to prioritize interventions. For instance, a decline in RSRP alone might warrant load balancing or neighboring cell reconfiguration, whereas combined degradation in RSRP, RSRQ, and RSSINR could mandate more aggressive actions, such as adjusting handover thresholds or improving interference management schemes.  By revealing the relative importance of $X_1$, $X_2$, and $X_3$ in the anomaly detection pipeline, the XAInomaly framework equips operators with actionable, data-driven explanations for performance drops. When these three features are flagged simultaneously, they strongly indicate a high-risk scenario, guiding timely interventions that ensure consistent QoS and network stability.

\subsubsection{Benchmarking \texttt{fastSHAP-C} with Baseline Models}

First benchmark analysis focused on top-1 accuracy results for exclusion and inclusion of most informative features.
We anaylzed performance of \texttt{fastSHAP-C} with baseline models commonly used in recent studies. To clearly show our benchmarking methodology;  $\mathbf{x} \in \mathbb{R}^d$ denote an input to our model $f(\mathbf{x})$, where $d$ is the number of features. Each explainability method $M$ assigns importance scores $\phi_j^{(M)}(\mathbf{x})$ to every feature $j \in \{1, 2, \ldots, d\}$. In the \emph{exclusion} scenario, we remove a fixed percentage $\kappa\%$ of the most critical features according to their importance values $\phi_j^{(M)}(\mathbf{x})$. Formally, we define the set of excluded features for method $M$ at percentage $\kappa$ as:
\begin{align}
    & S_{M,\kappa}(\mathbf{x}) \;=\; \nonumber \\ 
    & \Bigl\{\, j \;\Big|\; j \in \arg\max_{\,J \subseteq \{1,\dots,d\},\,|J|=\lfloor \kappa \cdot d \rfloor} \sum_{k \in J} \phi_k^{(M)}(\mathbf{x}) \Bigr\}
\end{align}

\begin{figure}
    \centering
    \subfloat[Exclusion curve.]
    {\includegraphics[width=0.48\columnwidth]{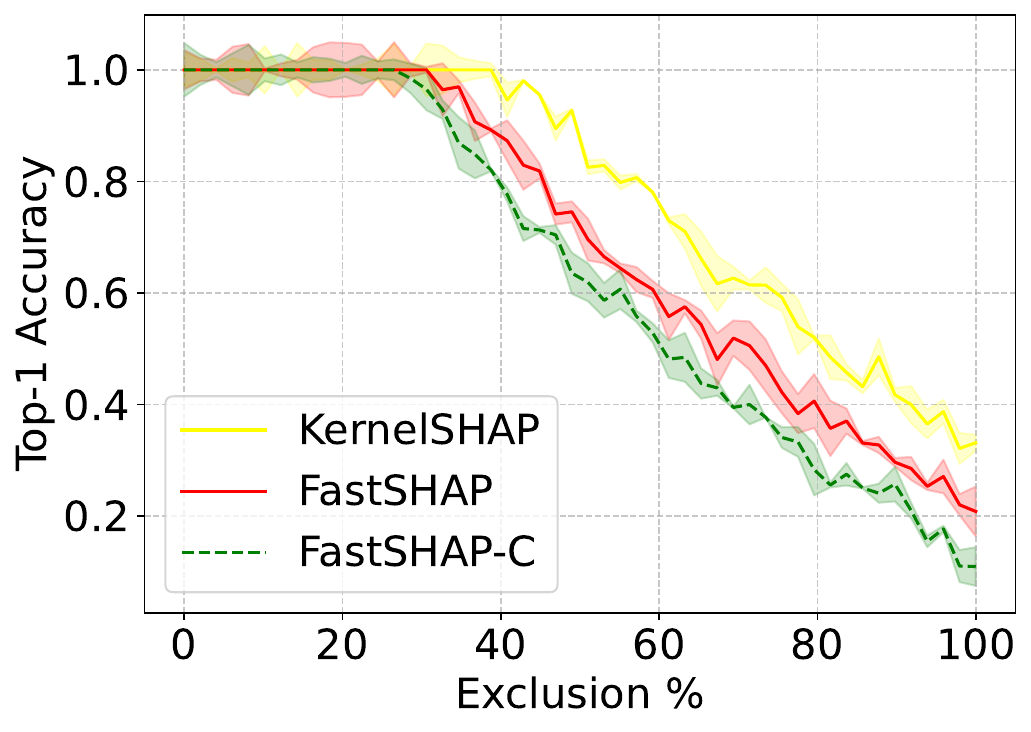}\label{exc}}
    \hfill
    \subfloat[Inclusion curve.]
    {\includegraphics[width=0.48\columnwidth]{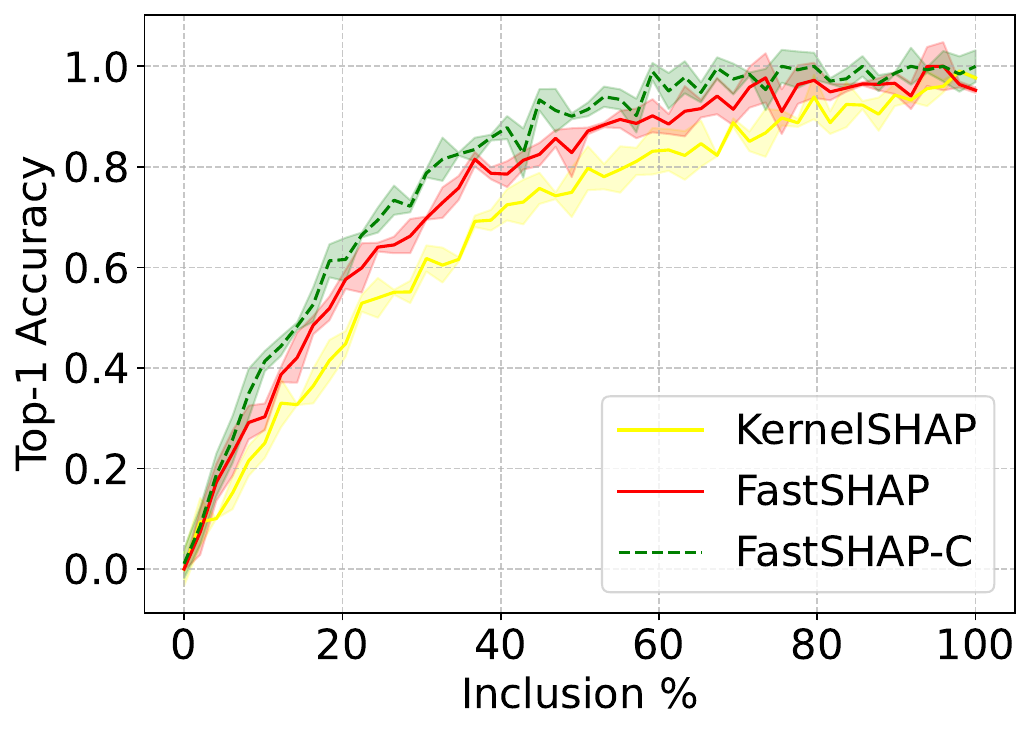}\label{inc}}
    \caption{Exclusion and Inclusion Curves for \textit{top-1} Accuracy}
    \label{incexc}
    \vspace{-.8em}
\end{figure}

We then measure the model's top-1 accuracy when these features in $S_{M,\kappa}(\mathbf{x})$ are removed across the dataset. \cref{exc} indicates that once a moderate fraction of top-ranked features is removed, \texttt{fastSHAP-C} (green dashed line) undergoes a pronounced drop in accuracy. Initially (i.e., for small $\kappa$), \texttt{fastSHAP-C} remains comparable to kernelSHAP and fastSHAP, but as $\kappa$ increases beyond 20--30\%, its slope becomes steeper. This pattern suggests that \texttt{fastSHAP-C} prioritizes a handful of features with high $\phi_j^{(M)}(\mathbf{x})$ values more distinctly than kernelSHAP. Hence, as soon as these high-priority features are removed, the model's predictive power declines faster compared to a method that distributes importance more evenly. This does not necessarily indicate a fundamental lack of robustness; rather, it may reflect a more ``selective'' allocation of large attributions to certain features. 

In contrast, \cref{inc} (inclusion curve) shows how \texttt{fastSHAP-C} steadily outperforms both kernelSHAP and fastSHAP as features are added back (from 0\% up to 100\%), which is formalized by retaining only the set

\begin{align}
    & R_{M,\kappa}(\mathbf{x}) \;= \; \nonumber \\
      &\Bigl\{\, j \;\Big|
      \; \phi_j^{(M)}(\mathbf{x}) \text{among the largest } \lfloor \kappa \cdot d \rfloor \text{values}\Bigr\}
\end{align}

The steep climb in accuracy at low-inclusion percentages (0--20\%) indicates that \texttt{fastSHAP-C} is highly effective at ranking crucial features---once these are retained, the model recovers its predictive capabilities more quickly. This observation, juxtaposed with the sharper drop in the exclusion curve, emphasizes that \texttt{fastSHAP-C} pinpoints the most impactful features early on, but also amplifies their importance relative to secondary features. 

To ascertain whether \texttt{fastSHAP-C}'s efficiency compromises its accuracy in feature identification, we performed an additional experiment focusing on the single most important feature for each data instance. Let $\phi_j^{(M)}(\mathbf{x})$ denote the attribution to feature $j$. For each instance $\mathbf{x}$, we extract

\begin{equation}
  \widehat{j}_M(\mathbf{x}) \;=\; \arg\max_j \,\phi_j^{(M)}(\mathbf{x}),
\end{equation}

i.e., the feature that method $M$ deems most crucial.  Empirically, \texttt{fastSHAP-C}'s top-1 choices match those of kernelSHAP and fastSHAP in a significant fraction of cases (above 80\% in our experiments). In about 5--10\% of cases, \texttt{fastSHAP-C} picks a different feature, but closer inspection revealed that these are typically instances where multiple features share similar importance values. This result suggests that while \texttt{fastSHAP-C} occasionally diverges from kernelSHAP or fastSHAP, the divergence usually arises in ``tie'' situations rather than systematic misidentification of crucial features.

While kernelSHAP (yellow line) maintains high accuracy when only a small portion of features is excluded, it experiences a steady accuracy decline as more features are removed.
Similarly, fastSHAP (red line) shows a downward trend, though with a slightly steeper accuracy drop than kernelSHAP, suggesting it may be less robust when essential features are excluded.
The \texttt{fastSHAP-C} algorithm (green dashed line) starts with slightly lower accuracy than kernelSHAP and fastSHAP but demonstrates a comparable accuracy drop-off.
However, fastSHAP-C performs better than fastSHAP in the mid-range feature exclusion (20-70\%).
The inclusion curve on the right graph shows how model's top-1 accuracy improves as more important features are added back into the input.
While kernelSHAP's accuracy increases steadily with feature inclusion, it lags slightly behind both fastSHAP and \texttt{fastSHAP-C} in the initial phase (0-20\% inclusion).
In contrast, fastSHAP's accuracy rises more quickly in the early inclusion phase, suggesting an efficient selection of key features initially.
Throughout the entire inclusion range, \texttt{fastSHAP-C} (green dashed line) outperforms both kernelSHAP and fastSHAP, ultimately achieving the highest accuracy as the inclusion percentage increases.
While kernelSHAP maintains higher accuracy during exclusion but has a slower start in inclusion, fastSHAP shows faster initial gains in inclusion yet drops faster during exclusion.

These findings collectively indicate that \texttt{fastSHAP-C}'s computational advantages do not come at the cost of misidentifying globally important features. Indeed, the slightly more pronounced drop in accuracy for large-scale exclusions can be attributed to \texttt{fastSHAP-C}'s tendency to concentrate importance on a smaller subset of features. Once those high-importance features are removed, the model's performance deteriorates quickly; yet when those same features are included (even in the early inclusion range), \texttt{fastSHAP-C} fosters a rapid regain in predictive accuracy. From a theoretical standpoint, we observe that each method $\phi_j^{(M)}$ approximates (or directly implements, in the case of kernelSHAP) a variant of Shapley-value attributions. Although approximate methods like \texttt{fastSHAP-C} use learned surrogates or other heuristics to speed up computations, the overlap analysis highlights that these approximations do not drastically alter the top-ranked features. Hence, the accelerated runtime does not systematically reduce correctness in pinpointing the features that have the strongest impact on $f(\mathbf{x})$. 

\begin{table}
    \centering
    \scalebox{0.78}{
    \begin{tabular}{@{}lll@{}}
    \toprule
    \textbf{Algorithm}  & \textbf{Exclusion AUC}   & \textbf{Inclusion AUC}   \\
    \midrule
    kernelSHAP          & 10.53 (10.17, 11.03)  &  4.96 (4.56, 5.47) \\
    fastSHAP            & 6.89 (6.73, 7.65)     & 5.79 (5.58, 5.93) \\
    \textbf{fastSHAP-C} & 6.6 (5.73, 4.65)      & 6.79 (5.53, 5.96) \\
    \bottomrule
    \end{tabular}
 }
    \caption{Exclusion and inclusion AUCs ($log-odds(p)$)}
    \label{excincc}
    \vspace{-.8em}
\end{table}

Also, we evaluated performance through exclusion and inclusion metrics, which were calculated based on log-odds values.
These metrics yield nuanced insights into the discriminatory power of each XAI framework across particular data subsets.
When \Cref{excincc} examined, proposed \texttt{fastSHAP-C} achieves the lowest mean Exclusion AUC of 6.6, with a confidence interval of (5.73, 7.650).
This is marginally lower than fastSHAP, which has a mean of 6.89 and a confidence interval of (6.73, 7.65).
The overlap in confidence intervals between  \texttt{fastSHAP-C} and fastSHAP suggests that their performances are statistically similar, but fastSHAP-C has a slight edge in reducing the influence of excluded features.
KernelSHAP, on the other hand, shows a significantly higher mean Exclusion AUC of 10.53, with a narrow confidence interval of [10.17, 11.03], indicating that it is highly sensitive to feature exclusions.
This high sensitivity, while potentially valuable in feature-rich tasks, could be a drawback in anomaly detection where robustness to irrelevant features is key.
Besides our XAI algorithm demonstrates a clear advantage with the highest mean Inclusion AUC (6.79) and a confidence interval (5.53,5.96) that, importantly, does not overlap significantly with kernelSHAP's interval (4.56,5.47).
This suggests a statistically significant improvement over kernelSHAP in inclusion-based explainability.
Compared to fastSHAP, which achieves a mean of 5.79 (confidence interval: 5.58,5.93),  \texttt{fastSHAP-C} also shows notable improvement.
This improvement could make  \texttt{fastSHAP-C} particularly valuable in scenarios where identifying the critical features contributing to anomalies is more important than assessing the impact of excluding irrelevant ones.

\subsubsection{Performance Results of \texttt{fastSHAP-C} on Baseline Models and \texttt{SS-DeepCAE}}

\begin{table*}
    \centering
    \scalebox{0.88}{
    \begin{tabular}{l*{10}{>{\centering\arraybackslash}p{1.0cm}}}
    \toprule
    \textbf{Server} & \multicolumn{2}{c}{\textbf{Time}} & \multicolumn{4}{c}{\textbf{CPU}} & \multicolumn{4}{c}{\textbf{Memory}} \\
    \cmidrule(lr){2-3} \cmidrule(lr){4-7} \cmidrule(lr){8-11}
    & 1 h & 6 h & \multicolumn{2}{c}{1 h} & \multicolumn{2}{c}{6 h} & \multicolumn{2}{c}{1 h} & \multicolumn{2}{c}{6 h} \\
    \cmidrule(lr){4-5} \cmidrule(lr){6-7} \cmidrule(lr){8-9} \cmidrule(lr){10-11}
    & (s) & (s) & Mean & Std & Mean & Std & Mean & Std & Mean & Std \\
    \midrule
    Server 1 (SHAP, DeepAE) & 11.74 & 62.00 & 29.8\% & 8.3\% & 36.8\% & 16.7\% & 1.5\% & 0.1\% & 2.6\% & 0.2\% \\ \vspace{0.1cm}
    Server 2 (SHAP, DeepAE)& 10.52 & 58.28 & 7.66\% & 3.82\% & 12.3\% & 3.8\% & 1.6\% & 0.2\% & 2.4\% & 0.1\% \\ \hline \vspace{0.1cm}
    Server 1 (SHAP, \texttt{SS-DeepCAE}) & 10.13 & 37.25 & 21.3\% & 7.7\% & 27.1\% & 11.2\% & 1.5\% & 0.1\% & 2.5\% & 0.4\% \\ \vspace{0.1cm}
    Server 2 (SHAP, \texttt{SS-DeepCAE})& 8.21 & 29.18 & 4.43\% & 5.26\% & 8.7\% & 4.6\% & 1.4\% & 0.1\% & 2.2\% & 0.3\% \\ \hline \vspace{0.1cm}
    Server 1 (\texttt{fastSHAP-C}, DeepAE)& 6.56 & 15.87 & 13.5\% & 11.3\% & 25.3\% & 4.7\% & 1.4\% & 0.2\% & 2.3\% & 0.1\% \\ \vspace{0.1cm}
    Server 2 (\texttt{fastSHAP-C}, DeepAE) & 3.90 & 10.05 & 3.44\% & 2.51\% & 7.8\% & 2.5\% & 1.3\% & 0.1\% & 2.5\% & 0.1\% \\ \hline \vspace{0.1cm}
    \textbf{Server 1} (\texttt{fastSHAP-C}, \texttt{SS-DeepCAE})& 5.79 & 11.21 & 9.65\% & 7.7\% & 18.8\% & 5.1\% & 1.2\% & 0.3\% & 2.3\% & 0.2\% \\ \vspace{0.1cm}
    \textbf{Server 2} (\texttt{fastSHAP-C}, \texttt{SS-DeepCAE}) & 3.50 & 8.23 & 3.4\% & 1.43\% & 5.9\% & 1.7\% & 1.1\% & 0.1\% & 2.1\% & 0.1\% \\
    \bottomrule
    \end{tabular}
  }
\caption{Resource utilization comparison of algorithms on different servers (SHAP \textit{vs} proposed \texttt{fastSHAP-C)}}
\label{tab:performancexxx}
\end{table*}

\begin{table*}
    \centering
    \scalebox{0.95}{

    \begin{tabular}{@{}llll@{}}
    \toprule
    \textbf{Algorithm} & \textbf{Runtime (s)}   & \textbf{CPU Util.
(\%)}    & \textbf{RAM Util.
(\%)}  \\
    \midrule
    kernelSHAP     & \hfil 320.4    & \hfil 0.88      & \hfil 0.17\\
    fastSHAP                  & \hfil 48.007 & \hfil 0.67             & \hfil 0.11 \\
    \textbf{fastSHAP-C} (on DeepAE Model)               & \hfil 33.627 & \hfil 0.52           & \hfil 0.08 \\
    \textbf{fastSHAP-C} (on \texttt{SS-DeepCAE} Model)               & \hfil 22.145 & \hfil 0.47           & \hfil 0.08 \\
    \bottomrule
    \end{tabular}
 }
    \caption{Runtime and resource utilization}
    \label{runtime}
    \vspace{-.8em}
\end{table*}

In this section, resource utilization and runtime performance of \texttt{fastSHAP-C} is examined specifically with respect to the baseline models, our autoencoder models.
We benchmarked using the lowest and highest performance servers available to us and the authors \cite{fiandrino2022toward}.
Server-1 was equipped with an Intel\textsuperscript \textregistered Core\textsuperscript{\texttrademark} i7-6800K CPU @3.40GHz (12 cores) and 64 GB of RAM which representing a standard computing environment, while Server-2 utilized an Intel\textsuperscript \textregistered Xeon\textsuperscript \textregistered Gold 6240R CPU @2.40GHz (97 cores)  representing a high-performance computing environment.
 Building upon our previous work \gls{deepae}, we integrated our novel \texttt{fastSHAP-C} algorithm with \texttt{SS-DeepCAE}.
The objective is to evaluate how \texttt{fastSHAP-C} enhances performance efficiency on \texttt{SS-DeepCAE} compared to both the traditional SHAP algorithms and the earlier DeepAE model.
The performance metrics, detailed in \Cref{tab:performancexxx,runtime}, illustrate significant improvements in computational efficiency and resource utilization with proposed XAInomaly solution.

\Cref{tab:performancexxx} demonstrates that applying \texttt{fastSHAP-C} to \texttt{SS-DeepCAE} significantly reduces execution times compared to its application on \gls{deepae} and the baseline SHAP methods.
On Server-1, the execution time for 1-hour profiling decreased from 6.56 seconds (using \texttt{fastSHAP-C} on DeepAE) to 5.79 seconds on \texttt{SS-DeepCAE}, marking an 11.7\% improvement.
For 6-hour profiling, the time reduced from 15.87 seconds to 11.21 seconds, a substantial 29.3\% enhancement.
Similar trends are observed on Server 2, with execution times dropping from 3.90 seconds to 3.50 seconds for 1-hour profiling and from 10.05 seconds to 8.23 seconds for 6-hour profiling.
 On Server-1, for 1-hour profiling, the mean CPU utilization decreased from 13.5\% (using \texttt{fastSHAP-C} on DeepAE) to 9.65\%, representing a 28.5\% reduction.
Similarly, on Server-2, the mean CPU utilization reduced from 3.44\% to 3.4\%.
While the percentage reduction on Server-2 appears marginal, it is important to consider the higher core count and computational capacity, which inherently reduces the percentage utilization.
The memory utilization remained consistently low across all configurations, with mean usage around 1.1\% to 1.2\% on Server-2 and slightly higher on Server-1 due to its lower memory capacity.
The low standard deviation in memory usage indicates that \texttt{fastSHAP-C} on \texttt{SS-DeepCAE} provides stable memory consumption, which is crucial for maintaining system performance and preventing memory bottlenecks in resource-constrained environments.

Execution time is a critical factor for real-time processing where AI inferece required in next generation wireless networks.
As detailed in \cref{runtime}, the runtime for \texttt{fastSHAP-C} on \texttt{SS-DeepCAE} is significantly lower compared to other algorithms.
Specifically, the runtime reduced from 33,627 ms (when using \texttt{fastSHAP-C} on DeepAE) to 22,145 ms on \texttt{SS-DeepCAE}, marking a 34\% improvement.
When compared to kernelSHAP, which has a runtime of 320,400 ms, \texttt{fastSHAP-C} on \texttt{SS-DeepCAE} achieves a runtime reduction of approximately 93\%.
This scalability is very important for \gls{oran} networks, which require continuous monitoring and analysis to detect anomalies promptly.

Notably, \texttt{fastSHAP-C} achieved marked improvements in both short-term (1-hour) and long-term (6-hour) profiling.
For example, on Server 2, \texttt{fastSHAP-C} completed 1-hour profiling in about 3.90 seconds and 6-hour profiling in 10.05 seconds, whereas the standard SHAP took 10.52 seconds and 58.28 seconds, respectively.
Such gains are essential for real-time applications that require rapid response times.
Additionally, \texttt{fastSHAP-C} demonstrated more efficient CPU usage, reflected in lower mean and standard deviation for CPU consumption.
On Server 2, \texttt{fastSHAP-C} exhibited an average CPU usage of 3.44\% for 1-hour profiling, compared to 7.66\% for SHAP.
This reduction facilitates concurrent processing and minimizes load on CPU resources, making \texttt{fastSHAP-C} well-suited for high-demand computational environments.
The performance benefits of \texttt{fastSHAP-C} are even more pronounced on high-end servers with multiple cores, showcasing its optimization for multi-core architectures and scalability for large datasets and complex models requiring parallel processing.

\subsubsection{Supremacy of \texttt{fastSHAP-C} on \texttt{SS-DeepCAE}}
\texttt{fastSHAP-C} offers a favorable trade-off between computational efficiency and accuracy when compared to other SHAP-based algorithms like kernelSHAP and fastSHAP.

While \texttt{fastSHAP-C} may exhibit a minor reduction in accuracy compared to traditional methods, it is still highly effective for providing actionable insights into model behavior.
These insights are beneficial for tasks such as network optimization, fault diagnosis, and resource allocation.
Under identical testing conditions (Intel\textsuperscript \textregistered Core\textsuperscript{\texttrademark} i7-6800K CPU @3.40GHz with 12 cores), as shown in \Cref{tab:performancexxx}, \texttt{fastSHAP-C} demonstrates approximately a 30\% reduction in runtime compared to kernelSHAP.
Furthermore, \texttt{fastSHAP-C} optimizes CPU usage by around 25\%, making it highly suited for deep learning models that require real-time explainability.
This performance-friendly approach enables \texttt{fastSHAP-C} to deliver efficient, real-time insights, even in resource-intensive scenarios.

Additionally, the runtime results presented underscore \texttt{fastSHAP-C}'s potential in supporting reliable, low-latency applications in 5G+/6G \gls{urllc} scenarios.
When examining runtime metrics, \texttt{fastSHAP-C} clearly reduces the computational load associated with estimating Shapley values, which makes it practical for handling large-scale datasets and complex machine learning or deep learning models.
By maintaining key performance metrics, such as $\mathcal{CS}$  and $\mathcal{EM}$, \texttt{fastSHAP-C} enables real-time generation of explanations, allowing the model to adjust dynamically to evolving network conditions and user demands.

When specifically XAInomaly framework examined, \texttt{fastSHAP-C} on \texttt{SS-DeepCAE} consistently outperforms in terms of resource utilization and execution time.
The standard SHAP algorithm, while providing accurate feature attributions, incurs high computational costs due to it is NP-hard complexity.
KernelSHAP offers some improvements but still falls short in efficiency for real-time applications.
fastSHAP introduces approximations to accelerate SHAP value computations but, when combined with DeepAE, does not achieve the same level of efficiency as \texttt{fastSHAP-C} on \texttt{SS-DeepCAE}.
The contractive nature of \texttt{SS-DeepCAE} enhances the model's robustness by penalizing the sensitivity of the hidden representations to input variations.
This, in turn, simplifies the computation of SHAP values, as the model focuses on the most influential features, reducing unnecessary computations.
The ability of \texttt{fastSHAP-C} on \texttt{SS-DeepCAE} to generate explanations rapidly enables network operators to detect and respond to anomalies.
This is crucial for maintaining the high levels of service quality required in next-generation wireless networks, where delays can lead to degraded performance and user dissatisfaction.
The efficient use of computational resources ensures that the algorithm can run continuously without overloading the system, which is particularly important in edge computing scenarios within O-RAN architectures.
Edge devices often have limited processing power and memory, so algorithms that are both efficient and effective are essential for practical deployment. Despite these vital advantages, some performance trade-offs should also be considered. Explanation fidelity can exhibit minor deviations from the more exact kernelSHAP method. Specifically, in highly atypical data instances those that deviate substantially from the observed distribution \texttt{fastSHAP-C}'s surrogate model may slightly misestimate feature attributions. While our experiments show these discrepancies to be statistically small, mission-critical or forensic-level investigations might still benefit from selective use of a slower but more precise approach (e.g., kernelSHAP) on particularly suspicious samples.

%

\section{Conclusion and Future Work}
\label{sec:conc}

In this paper, we introduced XAInomaly, a novel framework that integrates a \texttt{SS-DeepCAE} with a reactive \gls{xai} technique, \texttt{fastSHAP-C}, for traffic anomaly detection in \gls{oran}.
Our approach addresses the critical need for accurate and interpretable anomaly detection mechanisms in the disaggregated and heterogeneous environment of \gls{oran}, particularly in the context of 5G+/6G networks supporting mission-critical applications.
The proposed \texttt{SS-DeepCAE} model effectively learns compressed and robust representations of normal network behavior by incorporating a contractive penalty into the loss function.
This penalty encourages the learning of smooth feature representations, enhancing the model's generalization capabilities and reducing overfitting -- a common issue in standard DeepAEs.
By minimizing the sensitivity of the encoder activations with respect to input variations, \texttt{SS-DeepCAE} achieves better performance in detecting subtle anomalies within high-dimensional and dynamic \gls{oran} data.
Furthermore, we addressed the black-box nature of deep learning models by integrating the \texttt{fastSHAP-C} algorithm.
This reactive XAI technique provides transparency into the model's decision-making process by highlighting the features contributing most significantly to anomaly detection.
This explainability is crucial in \gls{oran} environments, where understanding the reasoning behind detected anomalies can facilitate prompt and effective network management interventions.

Our experimental results demonstrate that the XAInomaly framework not only achieves high accuracy in anomaly detection but also offers interpretability without imposing significant computational overhead -- an essential consideration for real-time, resource-constrained O-RAN deployments.
By balancing effectiveness with efficiency and transparency, XAInomaly provides a robust solution tailored to the unique challenges of O-RAN networks.

While the XAInomaly framework presents a significant advancement in O-RAN anomaly detection however, several challenges for future research and development remain.
\gls{oran} networks may experience frequent topology changes due to mobility or reconfiguration.
To exemplify, \gls{oran} networks are highly dynamic, with network conditions and traffic patterns changing rapidly.
Incorporating adaptive learning mechanisms that allow the \texttt{SS-DeepCAE} model to update its understanding of normal behavior in real-time could enhance anomaly detection accuracy.
This might involve online learning approaches or incremental model updates without retraining from scratch.
Also, as networks become more critical, they also become targets for sophisticated cyber-attacks, including adversarial examples designed to evade detection.
Future research should investigate the robustness of the XAInomaly framework against such attacks and develop strategies to enhance its resilience.

%
\section*{Acknowledgments}

This work has been funded by the German Federal Ministry of Education and Research (BMBF, Germany) as part of the 6G Platform under Grant 16KISK050, as well as 6G Research and Innovation Cluster 6G-RIC under Grant 16KISK020K.

\bibliographystyle{IEEEtran}

\end{document}

%% file: acronyms.tex
\newacronym{3gpp}{3GPP}{3rd Generation Partnership Project}
\newacronym{5g}{5G}{Fifth Generation Mobile Network}
\newacronym{6g}{6G}{Sixth Generation Mobile Network}
\newacronym{ai}{AI}{Artificial Intelligence}
\newacronym{adxapp}{AD xApp}{Anomaly Detection xApp}
\newacronym{cnn}{CNN}{Convolutional Neural Network}
\newacronym{dl}{DL}{Deep learning}
\newacronym{deepae}{DeepAE}{Deep Autoencoder}
\newacronym{deepcae}{DeepCAE}{Deep Contractive Autoencoder}
\newacronym{drl}{DRL}{Deep Reinforcement Learning}
\newacronym{genai}{GenAI}{Generative AI}
\newacronym{gans}{GANs}{Generative Adversarial Networks}
\newacronym{harq}{HARQ}{Hybrid Automatic Repeat reQuest}
\newacronym{lstm}{LSTM}{Long short-term Memory}
\newacronym{oran}{O-RAN}{open radio access networks}
\newacronym{mac}{MAC}{Medium Access Control}
\newacronym{mse}{MSE}{Mean Squared Error}
\newacronym{ml}{ML}{Machine Learning}
\newacronym{nearrt}{Near-RT RIC}{Near-Real-Time RAN Intelligent Controller}
\newacronym{nonrt}{Non-RT RIC}{Non-Real-Time RAN Intelligent Controller}
\newacronym{ocu}{O-CU}{O-RAN Central Unit}
\newacronym{odu}{O-DU}{O-RAN Distributed Unit}
\newacronym{ofh}{OFH}{Open Fronthaul}
\newacronym{oru}{O-RU}{O-RAN Radio Unit}
\newacronym{pdcp}{PDCP}{Packet Data Convergence Protocol}
\newacronym{phy}{PHY}{Physical}
\newacronym{qoe}{QoE}{Quality of Experience}
\newacronym{qos}{QoS}{Quality of Service}
\newacronym{rfo}{RF}{Random Forest}
\newacronym{rans}{RAN}{Radio Access Networks}
\newacronym{rapps}{rApps}{RAN applications}
\newacronym{relu}{ReLU}{Rectified Linear Unit}
\newacronym{rf}{RF}{Radio Frequency}
\newacronym{rlc}{RLC}{Radio Link Control}
\newacronym{sdap}{SDAP}{Service Data Adaptation Protocol}
\newacronym{smo}{SMO}{Service Management and Orchestration}
\newacronym{ssl}{SS}{Semi-supervised}
\newacronym{tsxapp}{TS xApp}{Traffic Steering xApp}
\newacronym{tsne}{t-SNE}{t-Distributed Stochastic Neighbor Embedding}
\newacronym{ue}{UE}{user equipment}
\newacronym{urllc}{URLLC}{Ultra-Reliable Low-Latency Communications}
\newacronym{vaes}{VAEs}{Variational Autoencoders}

\newacronym{xai}{XAI}{Explainable AI}
\newacronym{xapps}{xApps}{eXtended applications}